\documentclass[a4paper,fleqn,usenatbib]{mnras}
\usepackage[T1]{fontenc}
\usepackage{ae,aecompl}
\usepackage{graphicx}
\usepackage{amsmath}
\usepackage{amssymb}
\usepackage{txfonts}

\usepackage{float}
\usepackage{adjustbox}

\graphicspath{ {images/} }

\newcommand\spirou[1]{\textcolor{black}{#1}}
\newcommand\foca[1]{\textcolor{black}{#1}}

\newcommand{\ltsima}{$\; \buildrel < \over \sim \;$}
\newcommand{\lsim}{\lower.5ex\hbox{\ltsima}}
\newcommand{\gtsima}{$\; \buildrel > \over \sim \;$}
\newcommand{\gsim}{\lower.5ex\hbox{\gtsima}}
\newcommand{\bra}{\langle}
\newcommand{\ket}{\rangle}

\newcommand{\dd}{\mathrm{d}}

\newcommand{\ci}{\mathrm{i}}

\newcommand{\vecl}{\bmath{\ell}}

\newcommand{\dirac}{\delta_D}

\newcommand\BG[1]{\textcolor{black}{#1}}

\onecolumn

\title[Intrinsic sizes and shapes of galaxies]
{Intrinsic and extrinsic correlations of galaxy shapes and sizes in weak lensing data}
\author[B. Ghosh, R. Durrer, B.M. Sch{\"a}fer]
{Basundhara Ghosh$^1$, Ruth Durrer$^1$, Bj{\"o}rn Malte Sch{\"a}fer$^2$\thanks{e-mail: bjoern.malte.schaefer@uni-heidelberg.de}\\
$^1$D{\'e}partment de la Physique Th{\'e}orique, Universit{\'e} de Gen{\`e}ve, 24 quai Ernest Ansermet, 1211 Gen{\`e}ve, Switzerland\\
$^2$Zentrum f{\"u}r Astronomie der Universit{\"a}t Heidelberg, Astronomisches Rechen-Institut, Philosophenweg 12, 69120 Heidelberg, Germany
}

\begin{document}
\pagerange{\pageref{firstpage}--\pageref{lastpage}}
\pubyear{2021}
\maketitle
\label{firstpage}

\begin{abstract}
The subject of this paper \foca{is to build a physical model describing} shape and size correlations of galaxies due to weak gravitational lensing and due to direct tidal interaction of elliptical galaxies with gravitational fields sourced by the cosmic large-scale structure. Setting up a linear intrinsic alignment model for elliptical galaxies which parameterises the reaction of the galaxy to an external tidal shear field \foca{is controlled by} the velocity dispersion, we predict intrinsic correlations and cross-correlations with weak lensing for both shapes and sizes, juxtaposing both types of spectra with lensing. We quantify the observability of the intrinsic shape and size correlations and estimate with the Fisher-formalism how well the alignment parameter can be determined from the Euclid weak lensing survey. Specifically, we find a contamination of the weak lensing convergence spectra with an intrinsic size correlation amounting to up to 10\%  over a wide multipole range $\ell=100\ldots300$, with a corresponding cross-correlation exhibiting a sign change, similar to the cross-correlation between weak lensing shear and intrinsic shapes. A determination of the alignment parameter yields a precision of a few percent forecasted for Euclid, and we show that all shape and many size correlations should be measurable with Euclid. 
\end{abstract}


\section{introduction}\label{sect_intro}
Weak lensing has emerged as a powerful probe for investigating the cosmic large-scale structure \citep{mellier_probing_1999, bartelmann_weak_2001, amara_optimal_2007, bartelmann_gravitational_2010, kilbinger_cosmology_2015}, for testing gravitational theories and for constraining cosmological parameters. As gravitational lensing probes fluctuations in the gravitational potential directly \citep{kaiser_weak_1992, hu_weak_1999, hu_dark_2001, hu_dark_2002, bernstein_dark_2004, heavens_3d_2003,  heavens_measuring_2006, munshi_cosmology_2008, grassi_detecting_2014}, it depends on minimal assumptions and is fixed for a given gravitational theory. Correlations in the shapes of galaxies induced by weak lensing \citep{bernstein_shapes_2002, bernstein_comprehensive_2009} have been detected almost two decades ago, and by now lensing is recognised as a tool for investigating cosmological theories alongside the cosmic microwave background and galaxy clustering \citep{van_waerbeke_efficiency_1999, huterer_weak_2002, huterer_weak_2010,2014arXiv1401.0046M}. The last generation of surveys, most notably KiDS and DES \citep{Abbott:2017wau,Joudaki:2017zdt,Joudaki:2019pmv} have provided independent confirmation for the $\Lambda$CDM-model and support parameter determinations from the CMB, even though tensions between the two probes, most notably in the matter density $\Omega_m$ and $\sigma_8$ remain \citep{maccrann_cosmic_2014, Douspis:2018xlj}. The next generation of surveys, in particular Euclid \citep{Amendola:2016saw} and LSST~\citep{Abate:2012za} will probe cosmological models to almost fundamental limits of cosmic variance, but with decreasing statistical errors the control of systematical errors will become one of the central questions for data analysis, along with higher-order effects in the lensing signal related to evaluating the tidal shear fields along a geodesic \citep{ thomas_relativistic_2014}, effects of lensing on galaxy number counts~\citep{Ghosh:2018nsm} in galaxy-galaxy lensing correlation as well as non-Gaussian statistics of the lensing signal due to nonlinear structure formation and non-Gaussian contributions to the covariance \citep{jain_cosmological_1997, 2013arXiv1306.4684K, kayo_information_2013, munshi_tomography_2014}.

Among astrophysical contaminants of the weak lensing signal, intrinsic alignments \citep{jing_intrinsic_2002, mackey_theoretical_2002, heymans_weak_2004, altay_influence_2006, kirk_impact_2010, massey_origins_2013, kitching_limits_2016} are perhaps the most dramatic, leading to significant biases in the estimation of cosmological parameters, surpassing most likely baryonic corrections \citep{white_baryons_2004, semboloni_quantifying_2011}, \spirou{below multipoles of $\ell\leq 10^3$}. There are two primary models for the two dominant galaxy types for linking the apparent shapes to tidal gravitational fields in the large-scale structure \citep{dubinski_cosmological_1992}, which acts, due to long-ranged correlations, as the medium to reduce randomness and to correlate the measured ellipticities. The shapes of spiral galaxies are thought to be determined by the orientation of the angular momentum of the stellar disc \citep{catelan_intrinsic_2001, crittenden_spin-induced_2001, bailin_internal_2005}, and ultimately of the dark matter halo harbouring the stellar component. With this idea in mind, shape correlations are traced back to angular momentum correlations, which in turn would depend through tidal torquing as the angular momentum generated mechanism on the tidal shear fields. Tidal torquing models commonly predict ellipticity correlations on small scales at a level of at most 10\% of the weak lensing signal on multipoles above $\ell\simeq300$ for a survey like Euclid, many physical assumptions have been challenged, most notably the orientation of the disc relative to the host halo angular momentum, as well as an over-prediction of the correlation inherent to the torquing mechanism.

Elliptical galaxies, on the other hand, are thought to acquire shape correlations through direct interaction with the tidal shear field \citep{schneider_halo_2010, blazek_separating_2012, merkel_intrinsic_2013, PhysRevD.100.103506, tugendhat_angular_2018}: Second derivatives of the gravitational potential would give rise to an anisotropic deformation of the galaxy, in the principal directions of the tidal shear tensor. Interestingly, the reaction of a galaxy to the tidal shear field is determined by the inverse velocity dispersion $1/\sigma^2$ similar to lensing, where the relevant quantity is the gravitational potential in units of $c^2$. Tidal alignments of elliptical galaxies are thought to be present at intermediate angular scales of a few hundred in multipole $\ell$ for a survey like Euclid, with amplitudes being typically an order of magnitude smaller than that of the weak lensing effect. In parallel, alignment models using ideas from effective field theories provide parameterised relationships between tensors constructed from the cosmic density and velocity fields and can capture a wider range of alignment mechanisms and track them into the nonlinear regime \citep{vlah_eft_2019}, but perhaps with a less clear physical picture. There are indications that this in fact takes place in Nature, for instance in measurements of shape correlations in the local Universe \citep{brown_measurement_2002}, in shallow surveys \citep{lee_alignments_2007, chisari_cosmological_2013, pahwa_alignment_2016}, using stacking techniques or correlation techniques in deeper surveys \citep{hirata_galaxy-galaxy_2004, mandelbaum_wigglez_2011, chisari_intrinsic_2014} and correlation techniques in weak lensing surveys \citep{heavens_intrinsic_2000, heymans_weak_2003, kilbinger_dark-energy_2009, joachimi_constraints_2011, heymans_cfhtlens_2013, de_jong_kilo-degree_2013, jee_cosmic_2013, kilbinger_cfhtlens:_2013, schneider_galaxy_2013, kirk_cross-correlation_2015, joudaki_cfhtlens_2017, johnston_kids+gama:_2018}. Likewise, intrinsic alignment effects have been investigated in fluid-mechanical simulations of galaxy formation \citep[see for instance][]{tenneti_galaxy_2014, tenneti_intrinsic_2015, chisari_intrinsic_2015, debattista_internal_2015, chisari_redshift_2016, hilbert_intrinsic_2017, bate_when_2019}.

While intrinsic alignments refer to a physical change of the appearance of the galaxies \cite[for reviews, see][]{kiessling_galaxy_2015, joachimi_galaxy_2015, kirk_galaxy_2015, troxel_intrinsic_2015}, there is an analogous deformation effect on the shape of the light bundle emanating from a galaxy by gravitational lensing. To lowest order, both effects depend on tidal gravitational field which suggests that the effects must be correlated. The main difference is that while lensing shear comes from the gravitational tidal field integrated along the line of sight, intrinsic alignment is due to the local gravitational tidal field. Nevertheless, cross-correlations between the physical change in shape and the apparent change in shape are predicted to be nonzero for elliptical galaxies, and more precisely, should in fact be negative as galaxies align themselves radially with a large structure while lensing generates a tangential alignment. As a result, ellipticity correlations of galaxies is a sum of the conventional weak lensing (often referred to as $GG$), the intrinsic alignment (or $II$) and the cross-correlation between the two (called $GI$). Parameter estimation from weak lensing (\cite{casarini_non-linear_2011, capranico_intrinsic_2013, PhysRevD.100.103506}) as well as weak lensing mass reconstructions (\cite{fan_intrinsic_2007, chang_dark_2017}) would be affected by these intrinsic contributions, and can be taken care of by direct modelling or by self-calibration (\cite{troxel_self-calibration_2012, yao_effects_2017, yao_self-calibration_2018, yao_separating_2019, Pedersen:2019wfp}). In addition, intrinsic alignments can show up in cross correlation with the reconstructed CMB-lensing deflection field (\cite{hirata_cross-correlation_2004, hall_intrinsic_2014, chisari_contamination_2015, larsen_intrinsic_2016, merkel_imitating_2017}), and they might be usable as cosmological probes in their own right (\cite{pandya_can_2019, 2020ApJ...891L..42T}).

There should be analogous effects of the size of an elliptical galaxy due to tidal gravitational fields: In gravitational lensing the light bundle can be isotropically enlarged, i.e. changed in size while the shape is conserved: This nonzero convergence is caused by the trace of the tidal field, and determines to lowest order magnification as well, adding cosmological information (\cite{Huff:2011cq, takahashi_probability_2011}). Similarly, the size of an elliptical galaxy would physically change for a fixed velocity dispersion if the trace of the tidal field is nonzero\footnote{While in certain definitions the trace is subtracted in the tidal field, here the tidal field is simply $\partial_a\partial_b\Phi$ including the trace which is important as is it responsible for size changes.}, or equivalently, if it resides in an overdense or underdense region. An underdense region with density contrast $\delta < 0$ would source a gravitational potential $\Phi$ through the Poisson-equation $\Delta\Phi/c^2 = 3\Omega_m/(2\chi_H^2)\delta$, with the Hubble-distance $\chi_H = c/H_0$, such that the eigenvalues of $\partial_i\partial_j\Phi$ would be negative, stretching the galaxy to a physically larger size. Alternatively, one can argue that the change of volume (or area) is given by the Jacobian of the differential acceleration, i.e. of the tidal field, such that the perturbed volume is $V/V_0 = \mathrm{det}(\delta_{ab} + \partial_a\partial_b\Phi)$, implying that $\ln V-\ln V_0 = \ln\det(\delta_{ab} + \partial_a\partial_b\Phi) = \mathrm{tr}\ln(\delta_{ab} + \partial_a\partial_b\Phi) \simeq \mathrm{tr}(\partial_a\partial_b\Phi) = \Delta\Phi$ and consequently $V/V_0 = \exp(\Delta\Phi)$ and $(V-V_0)/V \simeq\Delta\Phi$. To what extent extrinsic and intrinsic size correlations can add to our understanding of cosmology has been investigated by \citet{heavens_combining_2013}.

The motivation of our paper is the study of these correlations between the sizes of elliptical galaxies as they would be predicted by a linear alignment model as a consequence of the trace $\Delta\Phi$ of the tidal tensor $\partial_a\partial_b\Phi$ being nonzero, as proposed by \citet{hirata_galaxy-galaxy_2004, hirata_intrinsic_2010}. These intrinsic size correlations would be generated in complete analogy to intrinsic shape correlations caused by the traceless part of the tidal, and would contaminate measurements of weak lensing convergence correlations (\cite{alsing_weak_2014}) in the same way as intrinsic shape correlations are a nuisance to the weak lensing shear. Alternatively, one can imagine these as a manifestation of ellipticity-density correlations (\cite{2002astro.ph..5512H}), only that density is mapped out by the galaxy size. After introducing tidal interactions of elliptical galaxies with their surrounding large-scale structure in Sect.~\ref{sect_tidal}, we compute shape correlations from direct tidal interaction and through gravitational lensing in Sect.~\ref{sect_spectra}. We quantify the information content of each of the correlations and the amount of covariance in Sect.~\ref{sect_fisher}, before discussing our results in Sect.~\ref{sect_summary}. In general we work in the context of a $w$CDM-cosmology with a constant equation of state value of $w$ close to $-1$, and standard values for the cosmological parameters, i.e. $\Omega_m = 0.3$, $\sigma_8 =  0.8$, $h = 0.7$ and $n_s = 0.96$, and a parameterised spectrum for nonlinearly evolving scales. We compute numerical results on the information content of size-correlations for the case of a tomographic weak lensing survey like Euclid's \cite{Amendola:2016saw}. Throughout the paper, summation convention is implied.

\section{tidal interactions of galaxies and gravitational lensing}\label{sect_tidal}
In a simplified way one can imagine elliptical galaxies as a stellar component in virial equilibrium with a velocity dispersion $\sigma^2$, filling the gravitational potential. \citet{piras_mass_2018} then argue that if the velocity dispersion is isotropic, one can invoke the Jeans-equation for stationary and static systems in order to relate density $\rho(r)$ and potential $\Phi(r)$,
\begin{equation}
\sigma^2\partial_r\ln\rho(r) = -\partial_r\Phi
\quad\rightarrow\quad
\rho(r) \propto \exp\left(-\frac{\Phi(r)}{\sigma^2}\right),
\label{eqn_jeans}
\end{equation}
reminiscent of the barometric formula. \spirou{Here, $r = 0$ is the centre of our galaxy where the density $\rho$ is highest and $\Phi$ has a minimum.} If the gravitational potential is distorted by external fields as the galaxy is not an isolated object, the equipotential contours get distorted correspondingly and the stellar component reacts and galaxy assumes a different shape.  We still assume that  $\Phi$ has a minimum at the center of the galaxy, $r = 0$. To lowest order, the change in shape takes place along the principal axes of the tidal tensor $\partial_a\partial_b\Phi$, which is defined as the tensor of second derivatives of the gravitational potential $\Phi$,
\begin{equation}
\Phi(r) \rightarrow \Phi(r) + \frac{1}{2}r_a r_b\partial_a\partial_b\Phi\:,
\label{eqn_tidal_taylor}
\end{equation}
leading to a distortion of the density of the stellar component. For weak tidal fields, the exponential can be Taylor-expanded to yield
\begin{equation}
\rho \propto 
\exp\left(-\frac{\Phi(r)}{\sigma^2}\right)\left[1-\frac{\partial_a\partial_b\Phi}{2\sigma^2}r_a r_b\right].
\end{equation}
For this perturbed stellar component one can compute the change of the second moments of the brightness distribution, where we ignore projection effects for a moment and use $\rho(r)$ for projected quantities,
\begin{equation}
\Delta q_{cd} = 
\int\dd^2r\:\rho(r)\: r_c r_d\: r_a r_b\times\frac{\partial_a\partial_b\Phi}{2\sigma^2} = S_{abcd}\:\Phi_{ab},   \qquad  \Phi_{ab} \equiv \partial_a\partial_b\Phi\,,
\label{eqn_2nd_moment}
\end{equation}
which bears a resemblance to the generalised Hooke-law $\Delta q_{cd} = S_{abcd}\:\Phi_{ab}$, relating the stresses $\Phi_{ab}$ to the observable strains $\Delta q_{cd}$, which suggests to think of $S_{abcd}$ as the susceptibility of a galaxy to change its shape or size under the influence of tidal gravitational fields. In the theory of elastic media one would then in fact use index symmetries to derive that there must be two material constants, similarly, in the theory of viscous fluids one defines two Lam{\'e}-viscosity coefficients (bulk and shear viscosity), so naturally the question arises whether the same constant of proportionality determines the size and the shape deformation as in the case of lensing. \spirou{As a consequence of the Jeans-equation~(\ref{eqn_jeans}), the relevant quantity for shape and size distortions is the tidal field normalised by the galaxy's velocity dispersion (\citet{camelio_origin_2015,piras_mass_2018}). With an extension of the virial law one can show that $R^2/\sigma^2$ is constant for a galaxy of size $R$ and velocity dispersion $\sigma^2$, as one would evaluate the Taylor-expansion~(\ref{eqn_tidal_taylor}) up to distances corresponding to $R$. Computing the second moment $\Delta q_{cd}$ for the ellipticity perturbation in eq.~(\ref{eqn_2nd_moment}) one obtains a scaling proportional to $R^2$, and using virial arguments again, with $M^{2/3}$, reproducing the observed behaviour of stronger alignments with increasing galaxy mass $M$.}

In our model we assume that the reaction of the galaxy to the tidal is instantaneous, which is an assumption that can be challenged: Adjustment to a new tidal field should take place on the free-fall time scale $t_\mathrm{ff} = 1/\sqrt{G\rho}$ with the total matter density $\rho$, that is typically a factor of $\Delta = 200$ higher than the background density $\Omega_m\rho_\mathrm{crit}$ with $\rho_\mathrm{crit} = 3H_0^2/(8\pi G)$. Substitution shows that the free fall time scale is only $\sqrt{8\pi/(3\Omega_m\Delta)}\simeq 0.37$ times shorter than the age of the Universe $1/H_0$, but because at least in linear structure formation tidal gravitational fields are close to constant in dark energy-cosmologies, the approximation might not be too bad. Of course in nonlinear structure formation, the time-scale of evolution would be much shorter and could give rise to an interesting time evolution of intrinsic alignments even for elliptical galaxies \citep{lee_nonlinear_2007, schafer_galactic_2012, schmitz_time_2018}. \spirou{Separated from the question of the time-dependence of the tidal interaction is whether the interaction can be the cause of shape distortions at all: \citep{camelio_origin_2015} have investigated this by considering the stellar distribution function in a perturbed potential and raise doubts whether alignments of the magnitude observed in ellipticity density correlations can be explained by a tidal alignment model: We would argue while acknowledging the difficulties of a self-consistent modelling that many details, for instance the tightly binding potential and the inconsistency introduced by working with a constant velocity dispersion and a cored potential can affect the results, as well as ignoring a dynamical change of the galaxy's own potential due to tidal interaction.}

After introducing polar coordinates, assuming spherical symmetry for the unperturbed galaxy and writing $r_0=r\cos\phi$ and $r_1=r\sin\phi$ for the vector components, the elasticity tensor is in our case given by
\begin{equation}
S_{abcd} = 
\frac{1}{2\sigma^2}\int\dd r\:r^5\rho(r)\int\dd\phi\:\cos^{4-(a+b+c+d)}\phi\sin^{a+b+c+d}\phi,
\end{equation}
has 16 entries and is fully symmetric under index exchange. \foca{Up to a numerical prefactor depending on the radial light distribution and on the velocity dispersion,} $S_{abcd}$ can only assume three different values, namely $S_{0000} = \int\dd\phi\:\cos^4\phi = S_{1111} = \int\dd\phi\:\sin^4\phi = 3\pi/4$, $S_{0001} = \int\dd\phi\:\cos^3\phi\sin\phi = S_{1110} = \int\dd\phi\:\cos\phi\sin^3\phi = 0$ and $S_{0011} = \int\dd\phi\:\cos^2\phi\sin^2\phi = \pi/4$.

\begin{figure}
\centering
\includegraphics[scale=0.45]{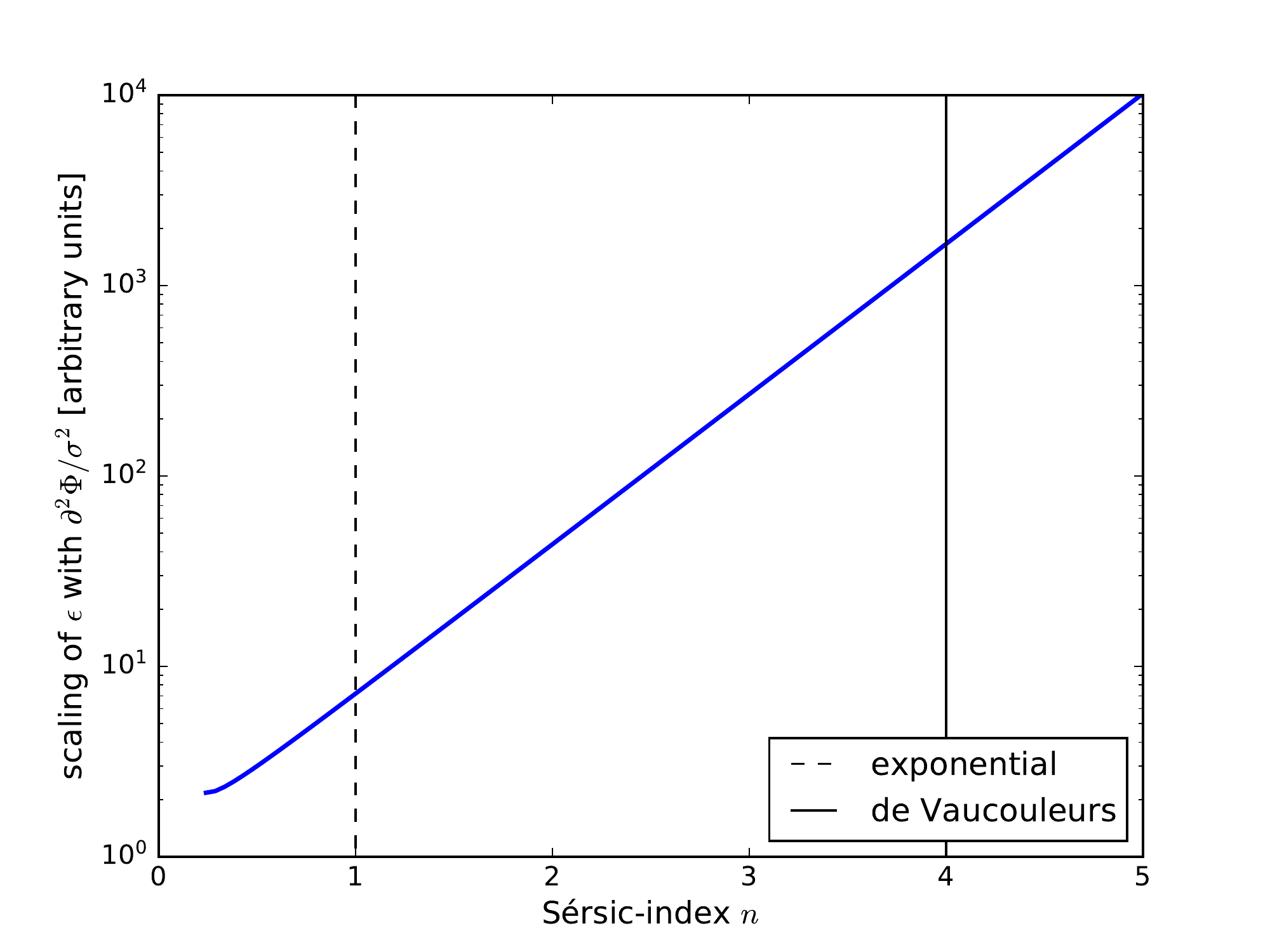}
\caption{Scaling of the relation between ellipticity $\epsilon$ and S{\'e}rsic-index $n$, for a given tidal gravitational field and a given velocity dispersion $\sigma^2$. As particular cases, the exponential profile for $n=1$ and the de Vaucouleurs-profile for $n=4$ are indicated by vertical lines.}
\label{fig_sersic_scaling}
\end{figure}

Let us introducing the four Pauli-matrices $\sigma^{(n)}_{ab}$ as the basis for the tidal $\partial_a\partial_b\Phi$,
\begin{equation}
\sigma^{(0)} = \left(
\begin{array}{cc}
+1 & 0 \\ 0 & +1
\end{array}
\right),~
\sigma^{(1)} = \left(
\begin{array}{cc}
+1 & 0 \\ 0 & -1
\end{array}
\right),~
\sigma^{(2)} = \left(
\begin{array}{cc}
0 & +1 \\ -1 & 0
\end{array}
\right),
\mathrm{~and~}
\sigma^{(3)} = \left(
\begin{array}{cc}
0 & +1 \\ +1 & 0
\end{array}
\right).
\end{equation}
Since $\sigma^{(2)}$ is anti-symmetric while the tidal tensor is symmetric as partial differentiations interchange, the component of $\Phi_{ab}$ in direction $\sigma^{(2)}$ vanishes. We now determine the change in size $s$ that is introduced by a tidal field $\Phi_{ab}\propto\sigma_{ab}^{(0)}$,
\begin{equation}
s = 
\frac{1}{2}\Delta q_{cd}\sigma^{(0)}_{cd} = 
\frac{1}{2}S_{abcd}\sigma^{(0)}_{cd}\sigma^{(0)}_{ab} = 
\frac{1}{2}\left(S_{0000} + S_{0011} + S_{1100} + S_{1111}\right) = 
\pi.
\end{equation}
The change in shape $\epsilon_+$ introduced by a tidal field $\Phi_{ab}\propto\sigma^{(1)}_{ab}$ is
\begin{equation}
\epsilon_+ = 
\frac{1}{2}\Delta q_{cd}\sigma^{(1)}_{cd} = 
\frac{1}{2}S_{abcd}\sigma^{(1)}_{cd}\sigma^{(1)}_{ab} = 
\frac{1}{2}\left(S_{0000} - S_{0011} - S_{1100} + S_{1111}\right) =
\frac{\pi}{2},
\end{equation}
while the change in shape $\epsilon_\times$ generated by the tidal field $\Phi_{ab}\propto\sigma^{(3)}_{ab}$ is given by
\begin{equation}
\epsilon_\times = 
\frac{1}{2}\Delta q_{cd}\sigma^{(3)}_{cd} =
\frac{1}{2}S_{abcd}\sigma^{(3)}_{cd}\sigma^{(3)}_{ab} = 
\frac{1}{2}\left(S_{0101} + S_{0110} + S_{1001} + S_{1010}\right) = 
\frac{\pi}{2}.
\end{equation}
The changes in shape, $\epsilon_{+,\times}$ are only half as large as the change in size, $s$, analogously to the weak lensing convergence with $\Delta\psi = 2\kappa$. With an assumption on the shape of the projected stellar density $\rho(r)$, for instance a S{\'e}rsic-profile \citep{sersic_influence_1963, graham_concise_2005},
\begin{equation}
\rho(r) \propto \exp\left(-b(n)\left[\left(\frac{r}{r_0}\right)^{1/n}+1\right]\right),
\label{eqn_sersic}
\end{equation}
it is possible to derive the scaling of ellipticity induced by the influence of a tidal gravitational field, dominantly with the size of the galaxy but also with the S{\'e}rsic-index $n$. In eqn.~(\ref{eqn_sersic}), $r_0$ is the scale radius of the stellar component, and $b(n)\simeq 2n - 1/3$, approximately \citep{de_vaucouleurs_recherches_1948}. Computing the relevant integral $\int\dd r\:r^5\rho(r)$ for a properly normalised density distribution $\int\dd^2r\:\rho(r) = 2\pi\int\dd r\:r\rho(r) = 1$ and using the definition of ellipticity $\epsilon$ as it would result from the second moments $q_{ab}$ of the normalised brightness distribution $I(r)$ which we equate to the stellar density $\rho(r)$,
\begin{equation}
\epsilon = \frac{q_{xx}-q_{yy}}{q_{xx}+q_{yy}} + 2\ci\frac{q_{xy}}{q_{xx}+q_{yy}},
\quad\mathrm{with}\quad
q_{ab} = \int\dd^2 r\:\rho(r)\:r_a r_b,
\end{equation}
where one recognises the size of the image in the denominator, $q_{xx} + q_{yy} = \int\dd^2r\:\rho(r)(x^2+y^2) = 2\pi\int\dd r\:r^3\rho(r)$, it is possible to show the scaling of the ellipticity to be
\begin{equation}
\epsilon \propto 
\left(\int_0^\infty\dd r\:r^5\rho(r)\right) \Bigg/ \left(\int_0^\infty\dd r\:r^3\rho(r)\right) = 
r_0^2 \times \int_{-b}^\infty\dd x\:\left(\frac{x}{b}+1\right)^{6n-1}\exp(-x) \Bigg/ \int_{-b}^\infty\dd x\:\left(\frac{x}{b}+1\right)^{4n-1}\exp(-x).
\end{equation}
Technically, we obtained this result after substitution $x = b\left[(r/r_0)^{1/n}-1\right]$, where the ratio of integrals has in general only a numerical solution and shows the dependence of the susceptibility to shape change due to tidal forces caused by the distribution of the stars inside the galaxy. The dominant scaling of ellipticity with the size $r_0^2$ of the galaxy is dimensionally consistent with the linear tidal model $q_{ab} = S_{abcd}\:\Phi_{cd}$. The results are shown in Fig.~\ref{fig_sersic_scaling}, which indicates a strong scaling of the alignment parameter with increasing S{\'e}rsic-index $n$, where we should note that we consider the S{\'e}rsic-profile as a reasonably simple model for the stellar distribution, which is not consistent with a constant velocity dispersion $\sigma^2$, and neither a gravitating self-consistent solution. Rather, it is supposed to illustrate that the internal dynamics of an elliptical galaxy can impact on the magnitude of tidal alignment and that not all elliptical galaxies should have the same alignment parameter if their S{\'e}rsic-index varies. \spirou{We would like to mention here that effects like isophotal twisting are neglected in our model, and are discussed in literature such as \cite{Singh:2014kla}. Effectively, isophotal twisting would correspond to different measurements of ellipticity at different radii of an elliptical galaxy and is possibly induced by a complex velocity structure of the elliptical galaxy: How these features could be related to tidal interaction and anisotropic accretion is not well understood and subject of debate.}

It is straightforward to show that the distortion modes are all independent for the linear model, i.e. tidal fields $\propto\sigma^{(m)}_{ab}$ will never source distortion modes $\propto\sigma^{(n)}_{cd}$ with $m\neq n$. For making the influence of the tidal field on the galaxy size more specific, we compute the change in size $s$ explicitly as the second moment of the brightness distribution for the isotropic case,
\begin{equation}
s = 
\frac{1}{2\sigma^2}\int\dd^2r\:r^2\rho(r)\left[\frac{1}{2}\partial_a\partial_b\Phi\: r_ar_b\right] =
\frac{1}{2\sigma^2}\int\dd^2r\:r^2\rho(r)\left[\frac{1}{4}\Delta\Phi\:r^2\right] = 
\frac{\pi}{\sigma^2}\int r^5\dd r\:\rho(r)\frac{\Delta\Phi}{4} \propto \frac{\pi}{2}\Delta\Phi
\end{equation}
such that the change in size comes out proportional to the trace $\Delta\Phi$ of the tidal field. 

Defining the alignment parameter $D_{IA}$ as the constant of proportionality between ellipticity tensor $\epsilon_{ab} =q_{ab}/(q_{xx}+q_{yy})$ and the tidal gravitational field $\partial_a\partial_b\Phi$,
\begin{equation}
\epsilon_{ab} \simeq \frac{1}{2}\left(\frac{c}{\sigma}\right)^2\frac{\int\mathrm{d}^2r\:\rho\:r^4}{\int\mathrm{d}^2r\:\rho\:r^2}\times\partial_a\partial_b\:\frac{\Phi}{c^2}  ~\equiv D_{IA}\times\partial_a\partial_b\:\frac{\Phi}{c^2},
\label{eqn_scaling}
\end{equation}
determines the unit to be comoving $(\mathrm{Mpc}/h)^2$ (we compute the derivatives with respect to comoving coordinates, and relate it to the CDM-spectrum $P(k)$ as a function of comoving wave number $k$. We also carry out the Limber-projection with a line-of-sight integration in comoving coordinates for consistency). The same argument is valid for the relationship between size $s$ and the trace $\Delta\Phi$. In order to complete the analogy with lensing, we work with the dimensionless potential $\Phi/c^2$ as the quantity that appears as the metric perturbation.

The mass scaling of the alignment parameter $D_{IA}$ requires the combination of a number of arguments evolving around the virial law and galaxy scaling relations: Assuming the weighted volume element $\mathrm{d}^2r\:\rho$ to be scalar and assuming a scaling of the size $r$ of the galaxy proportional to $M^{1/3}$, as well as the virial relation $\sigma^2 = GM/r$, in order to relate the specific kinetic and potential energies with the size $r$ and mass $M$ of the galaxy implies a scaling of the ratio of the two integrals in (\ref{eqn_scaling}) $\propto M^{2/3}$, which is exactly cancelled by the inverse velocity dispersion, $\sigma^2\propto r^2 \propto M^{2/3}$, such that the entire alignment parameter $D_{IA}$ should be independent of the mass, as long as the virial argument and the scaling of galaxy size with mass is valid.

Concerning the actual numerical value of $D_{IA}$ one can set up an interesting argument by means of the comoving Poisson-equation $\Delta\Phi/c^2 = 3\Omega_m/(2\chi_H^2)\:\delta$: In fact, replacing the derivatives $\partial_i\partial_j\Phi$ with $\Delta\Phi$ which is sufficient for scaling arguments, then yields the expression
\begin{equation}
\epsilon \sim \frac{1}{2}\left(\frac{c}{\sigma}\right)^2\frac{\int\mathrm{d}^2\rho\:r^4}{\int\mathrm{d}^2r\rho\:r^2}\times\frac{3\Omega_m}{2\chi_H^2}\delta.
\end{equation}
For a typical density perturbation of order $\delta \simeq 1$ one obtains an intrinsic size or ellipticity perturbation of $\epsilon\simeq10^{-5}$ if the velocity dispersion has a value of $\sigma=10^5\mathrm{m}/\mathrm{s}$, yielding $c/\sigma\simeq 10^7$ and if the ratio of the integrals in eqn.~(\ref{eqn_scaling}) is set to a value of $10^{-5}~(\mathrm{Mpc}/h)^2$, which we call the alignment parameter in the remainder of the paper. The factor $3\Omega_m/(2\chi_H^2)$ is typically  of order $\simeq10^{-7}(\mathrm{Mpc}/h)^{-2}$. The alignment parameter physically depends on the area of the galaxy. Assuming an exponential S{\'e}rsic-profile with $n=1$  then implies that the scale length of a galaxy of mass $10^{12}M_\odot/h$ is a few $kpc/h$, which coincides roughly with the numerical value of $10^{-5}~(\mathrm{Mpc}/h)^2$ chosen in our study.  A comparison with other values of $D_{IA}$ will follow in Sect.~\ref{sect_spectra}.

From our derivation it becomes apparent that the reaction of the galaxy to a tidal gravitational field in terms of  shape and size distortions is governed by the same parameter. Shape and size distortions are merely different and mutually orthogonal modes of the second moments of the brightness distribution, which our linear model relates directly to the tidal gravitational field. This situation is completely analogous with gravitational lensing.

Deviations from this simple scaling of $D_{IA}$ with mass at lowest order can be due to a number of effects: There can be systematic trends of the virial relationship with mass, deviating from a pure power law. The smoothing of the tidal field on the mass scale will certainly have an influence on the measured alignment parameter as it impacts the spectra on small scales, as well as the Abel-integration which relates the 3-dimensional light distribution to the 2-dimensional one, and finally the scale radius $r_0$ in the S{\'e}rsic-profile family might deviate from the scaling $\propto M^{1/3}$, similar to the concentration parameter in the NFW-profile family. It has also been argued that more massive galaxies tend to have larger  S{\'e}rsic-index \citep{10.1093/mnras/stu2467}, leading to a mass-dependence of the alignment parameter. We  point out that residual scaling properties of the alignment parameter with mass can be significant, if galaxies over a wide range of masses are observed, since the steepness of the Press-Schechter function biases  averages towards the value of $D_{IA}$ at low masses.

It is interesting to note that shape and size distortion caused by a single perturbation are comparable for intrinsic alignments and for gravitational lensing, but lensing, as an integrated effect, dominates ultimately. Repeating the above argument would lead to expressing the weak lensing shear given by
\begin{equation}
\gamma = \frac{3\Omega_m}{2\chi_H^2}\int_0^{\chi_H}\mathrm{d}\chi\:\frac{\chi_H-\chi}{\chi_H}\chi\:\delta
\end{equation}
in the simplest case, with a single source distance at $\chi_H$. Placing a single lens with the longitudinal extension of $\Delta\chi = 1~\mathrm{Mpc}/h$ and density $\delta$ at $\chi = \chi_H/2$ yields $\kappa\simeq10^{-4}$. Combining all shape and size distortions in an uncorrelated random walk with $\chi_H/\Delta\chi$ steps amplifies the signal by a factor of $\sqrt{\chi_H/\Delta\chi}$, which brings the lensing  signal to values between $10^{-3}$ and $10^{-2}$, which one typically cites as a weak shear distortion.

With many galaxies in a tomographic bin $A$ with a suitable, normalised redshift distribution $p_A(z)\dd z$ one can define the line of sight-averaged ellipticity from second angular derivatives of the weighted projection of the potential $\Phi$:
\begin{equation}
\varphi_{A,ab} = \partial_a\partial_b\varphi_A
\quad\mathrm{with}\quad
\varphi_A = D_{IA}\int\dd\chi\:p_A(z(\chi))\frac{\dd z}{\dd\chi}\frac{1}{\chi^2}\frac{D_+(a)}{a}\:\frac{\Phi}{c^2} = \int\dd\chi\:W_{\varphi,A}(\chi)\:\frac{\Phi}{c^2},
\label{eqn_ia_los}
\end{equation}
with the Hubble-function $H(\chi)/c = \dd z/\dd\chi$ which originates from the transformation of the redshift distribution, and the growth rate $D_+(a)/a$ of gravitational potentials, and the alignment parameter $D_{IA}$, which encapsulates the proportionality between tidal field and physical shape and size change. The line of sight-weighting function $W_{\varphi,A}$ of bin $A$ is defined by the last equals sign. The parameter $D_{IA}$ reflects the brightness distribution of a galaxy through its second moments and scales inversely with the velocity dispersion $\sigma^2$. Because linear intrinsic alignments have opposite signs compared to gravitational lensing in the same gravitational potential, we choose a negative value for the alignment parameter $D_{IA}$ in order to not having to carry through minus-signs explicitly. This is due to the fact that an overdense region enlarges the image of a galaxy in lensing but compresses a galaxy physically.

Modelling the statistics of the intrinsic alignment effects from a Gaussian random field as we do in equation~(\ref{eqn_ia_los}) subsequently ignores that the galaxy shapes and sizes provide a measurement of the tidal field restricted to peak regions of the large-scale structure, which influences the statistics of tidal fields \citep{peacock_statistics_1985, schafer_galactic_2012}. On the other hand, correlations between tidal fields and characterisations of the environment, for instance in terms of the eigenvalues of the tidal shear tensor \citep{forero-romero_cosmic_2014, reischke_environmental_2018}, should apply directly to correlations of shear or size as observables. A related point would be the introduction of a density weighting of the tidal gravitational fields that give rise to the intrinsic shape and size correlations: Those weightings are straightforward to construct as they involve a convolution between the tidal field and density spectra, if one restricts oneself to a reasonably simple multiplicative weighting. This density weighting would associate terms analogous to source-source and source-lens clustering known from gravitational lensing with those arising in alignment models, but would necessitate new parameters and would introduce non-Gaussian statistical properties. In a certain sense, effective field theory models of intrinsic alignments pursue exactly this route \citep{PhysRevD.100.103506}. Phenomenologically, enhancing the ellipticity correlation function with a linear and deterministic biasing term would be possible, too, but with an ambiguous physical interpretation: While it is clear that the observed ellipticity correlations arise from tidal field correlation with a modulation mediated by the galaxy density, only the relation between galaxy size and galaxy density is given by the ambient matter density, as a consequence of the Poisson-relation $s \sim \Delta\Phi \sim \delta$.

The angular derivatives $\partial_a$ are related to the spatial derivatives $\partial_x$ through $\partial_a = \chi\partial_x$, with $x=\theta\chi$ in the small-angle approximation. From that, one can recover the ellipticity components $\epsilon_{+,A}$ and $\epsilon_{\times,A}$ as well as the size $s_A$ from a decomposition of the tensor $\varphi_{A,ab}$ with the Pauli-matrices $\sigma_{ab}^{(n)}$,
\begin{equation}
\varphi_{A,ab} = s_A\sigma^{(0)}_{ab} + \epsilon_{+,A}\sigma^{(1)}_{ab} + \epsilon_{\times,A}\sigma^{(3)}_{ab},
\end{equation}
where these three components are sufficient because of the symmetry $\varphi_{A,ab} = \varphi_{A,ba}$. Using two properties of the Pauli-matrices $\sigma_{ab}^{(n)}$, namely $\sigma_{ab}^{(l)}\sigma_{bc}^{(m)} = \delta_{lm}\sigma^{(0)}_{ac} + \epsilon_{lmn}\sigma^{(n)}_{ac}$, and their tracelessness $\sigma^{(m)}_{aa} = 0$, it is possible to invert the last relation and to obtain the expansion coefficients,
\begin{equation}
s_A = \frac{1}{2}\varphi_{A,ab}\sigma^{(0)}_{ab},
\quad
\epsilon_{+,A} = \frac{1}{2}\varphi_{A,ab}\sigma^{(1)}_{ab},
\mathrm{~and}\quad
\epsilon_{\times,A} = \frac{1}{2}\varphi_{A,ab}\sigma^{(3)}_{ab}.
\end{equation}

The approach above is motivated by the weak lensing shear $\gamma$ in some bin $B$, which results from the tensor $\psi_{B,ab}$ containing the second derivatives of the weak lensing potential $\psi_B$,
\begin{equation}
\psi_{B,ab} = \partial_a\partial_b\psi_B
\quad\mathrm{with}\quad
\psi_B = 
2\int\dd\chi\:\frac{G_B(\chi)}{\chi}\frac{D_+(a)}{a}\frac{\Phi}{c^2} = 
\int\dd\chi\:W_{\psi,B}(\chi)\frac{\Phi}{c^2},
\end{equation}
with the lensing efficiency
\begin{equation}
G_B(\chi) = \int_{\mathrm{max}(\chi,\chi_B)}^{\chi_{B+1}}\dd\chi^\prime\:p_B(\chi^\prime)\frac{\dd z}{\dd\chi^\prime}\left(1-\frac{\chi}{\chi^\prime}\right).
\end{equation}
It is interesting to note that the effect of convergence and shear are fully analogous to the changes in size and shape due to direct tidal interaction, up to some interesting details: A light bundle, consisting of photons as relativistic test particles for the gravitational potential, is deflected twice as strongly compared to non-relativistic test particles such as the stars inside an elliptical galaxies, and the constant of proportionality that makes the gravitational potential dimensionless is $c^2$ in lensing instead of $\sigma^2$ for the intrinsic alignments. Finally, the lensing kernel $G_B/\chi$ is  non-zero not only inside the bin $B$ under consideration but the integral extends from $\chi=0$ to the outer rim of bin $B$, $\chi_{B+1}$. We  compute both lensing and intrinsic alignments from the dimensionless potential $\Phi$ give in units of $c^2$ and we use a numerical value for the alignment parameter scaled by $c^2/\sigma^2$. Again, there is an analogous decomposition
\begin{equation}
\psi_{B,ab} = \kappa_B\sigma^{(0)}_{ab} + \gamma_{+,B}\sigma^{(1)}_{ab} +\gamma_{\times,B}\sigma^{(3)}_{ab}
\end{equation}
with the analogous inversion,
\begin{equation}
\kappa_B = \frac{1}{2}\psi_{B,ab}\sigma^{(0)}_{ab},
\quad
\gamma_{+,B} = \frac{1}{2}\psi_{B,ab}\sigma^{(1)}_{ab},
\mathrm{~and}\quad
\gamma_{\times,B} = \frac{1}{2}\psi_{B,ab}\sigma^{(3)}_{ab}.
\end{equation}
The intrinsic size field provides a measure of the projected density in the same way as the weak lensing convergence $\kappa$, but with a different weighting function:
\begin{equation}
s = 
\frac{1}{2}\varphi_{ab}\sigma^{(0)}_{ab} = 
\frac{D_{IA}}{2}\sigma^{(0)}_{ab}\partial_a\partial_b\int\dd\chi\: p(\chi)\frac{1}{\chi^2}\frac{D_+(a)}{a}\frac{\Phi}{c^2} = 
\frac{D_{IA}}{2}\int\dd\chi\:p(\chi)\frac{D_+(a)}{a}\frac{\Delta\Phi}{c^2} = 
\frac{3\Omega_m}{4\chi_H^2}D_{IA}\int\dd\chi\:p(\chi)\frac{D_+(a)}{a}\delta\,.
\end{equation}
We have substituted the Poisson-equation $\Delta\Phi/c^2 = 3\Omega_m/(2\chi_H^2)\delta$, using $\partial_a = \chi\partial_x$ for the derivatives, and approximated the full Laplacian by the one containing the derivatives perpendicular to the line of sight, as well as the Hubble-distance $\chi_H = c/H_0$. Again, one recognises a factor of two between the gravitational acceleration of photons in gravitational lensing and non-relativistic particles as in our case of stars inside an elliptical galaxy. As discussed before, an actual measurement of the mean size $s$ of the galaxies into a certain direction would in addition be weighted with a biasing factor because the tidal field is only measurable at positions where galaxies exist: While the inclusion of a reasonably simple linear and deterministic biasing model is certainly possible and straightforward, we ignore this here for simplicity. 

This implies that the statistics of all modes of the shape and size field can be described by spectra of the source fields, which in turn are given by a Limber-projection. Specifically, the spectrum of $\varphi_{A,ab}$ reads
\begin{equation}
\bra\varphi_{A,ab}(\bmath\ell)\varphi_{B,cd}^*(\bmath\ell^\prime)\ket = 
(2\pi)^2\dirac(\bmath\ell-\bmath\ell^\prime)\:C^{\varphi_A\varphi_B}_{abcd}(\ell)
\quad\mathrm{with}\quad
C^{\varphi_A\varphi_B}_{abcd}(\ell) = 
\ell_a\ell_b\ell_c\ell_d\:\int\frac{\dd\chi}{\chi^2}\:W_{\varphi,A}(\chi)W_{\varphi,B}(\chi)\:P_{\Phi\Phi}(k = \ell/\chi),
\end{equation}
similarly, one obtains for the  field $\psi_{B,ab}$,
\begin{equation}
\bra\psi_{A,ab}(\bmath\ell)\psi_{B,cd}^*(\bmath\ell^\prime)\ket = 
(2\pi)^2\dirac(\bmath\ell-\bmath\ell^\prime)\:C^{\psi_A\psi_B}_{abcd}(\ell)
\quad\mathrm{with}\quad
C^{\psi_A\psi_B}_{abcd}(\ell) = 
\ell_a\ell_b\ell_c\ell_d\:\int\frac{\dd\chi}{\chi^2}\:W_{\psi,A}(\chi)W_{\psi,B}(\chi)\:P_{\Phi\Phi}(k = \ell/\chi),
\end{equation}
and finally for their cross-correlation,
\begin{equation}
\bra\varphi_{A,ab}(\bmath\ell)\psi_{B,cd}^*(\bmath\ell^\prime)\ket =
(2\pi)^2\dirac(\bmath\ell-\bmath\ell^\prime)\:C^{\varphi_A\psi_B}_{abcd}(\ell)
\quad\mathrm{with}\quad
C^{\varphi_A\psi_B}_{abcd}(\ell) =
\ell_a\ell_b\ell_c\ell_d\:\int\frac{\dd\chi}{\chi^2}\:W_{\varphi,A}(\chi)W_{\psi,B}(\chi)\:P_{\Phi\Phi}(k = \ell/\chi),
\end{equation}
\spirou{where the four powers of $\ell$ arise through the double differentiation of the potentials $\phi$ and $\psi$ to link them to tidal fields, which are subsequently squared in computing the spectrum.} In general, all lensing effects originating from a tidal gravitational field will have the opposite sign than the intrinsic tidal alignment, which causes the cross-correlation between lensing and intrinsic alignments to have a negative sign. This is taken care of numerically by choosing a negative value for the alignment parameter $D_{IA}$, which does not affect the auto-correlations: Those are proportional to $D_{IA}^2$ and therefore positive. In analogy we define the angular spectra $C^{\varphi_A\varphi_B}(\ell)$, $C^{\psi_A\psi_B}(\ell)$ and $C^{\varphi_A\psi_B}(\ell)$ of the potentials $\varphi_A$ and  $\psi_B$. For the spectrum of the gravitational potential we use a linear spectrum of the form $P_{\Phi\Phi}(k)\propto k^{n_s-4}T^2(k)$ with a transfer function $T(k)$ and a nonlinear extension on small scales \citep{cooray_power_2001, huterer_calibrating_2005}, normalised to $\sigma_8$, but assume Gaussian statistics throughout. We apply a smoothing on a scale defined through $M = 4\pi/3\:\Omega_m\rho_\mathrm{crit}R^3$, $\rho_\mathrm{crit} = 3H_0^2/(8\pi G)$, 
\begin{equation}
\Phi(k) \rightarrow \Phi(k)\exp\left(-\frac{(kR)^2}{2}\right),
\end{equation}
to the potential used for intrinsic alignments, \spirou{where we set the mass scale to be that of a small elliptical galaxy, $M = 10^{12}M_\odot/h$: In doing this we can control how close size- and shape-correlations trace the tidal shear field, and select the relevant long-wavelength modes. We set value of the velocity dispersion $\sigma^2$ to be consistent with the mass scale according to the assumed virial equilibrium.}

\section{Angular spectra of galaxy shapes and sizes}\label{sect_spectra}
The prefactors $\ell_a\ell_b$ appearing in the expressions for the spectra of the projected tidal shears can be compactly written by introducing polar coordinates, $\ell_0 = \ell\cos\phi$ and $\ell_1 = \ell\sin\phi$. Then,
\begin{equation}
\ell_a\ell_b = 
\frac{\ell^2}{2}\left(\sigma^{(0)}_{ab} + (\cos^2\phi-\sin^2\phi)\sigma^{(1)}_{ab} + 2\sin\phi\cos\phi\sigma^{(3)}_{ab}\right) = 
\frac{\ell^2}{2}\left(\sigma^{(0)}_{ab} + \cos(2\phi)\sigma^{(1)}_{ab} + \sin(2\phi)\sigma^{(3)}_{ab}\right),
\end{equation}
recovering the fact that the phase angle rotates twice as fast as the coordinate system. We are going to make the choice $\phi = 0$ by a suitable rotation of the coordinate frame, such that there are no contractions with $\sigma^{(3)}_{ab}$, and correspondingly vanishing $\gamma_\times$ or $\epsilon_\times$. This corresponds effectively to the computation of $E$- and $B$-modes of the shear field and of the ellipticity field, with
\begin{align}
e(\vecl) = &\hphantom{-}\cos(2\phi)\gamma_+(\vecl) + \sin(2\phi)\gamma_\times(\vecl),\\
b(\vecl) = &-\sin(2\phi)\gamma_+(\vecl) + \cos(2\phi)\gamma_\times(\vecl),
\end{align}
where in our model there are no $B$-modes due to the index exchange symmetry. \spirou{This is in contrast to the predictions of tidal torquing models for spiral galaxies, where an ellipticity $B$-mode comparable to the ellipticity $E$-mode appears very naturally. For Gaussian tidal gravitational fields, one would not expect cross-correlations between the shapes of spirals and ellipticals, if in fact the alignment of spirals is given by the quadratic tidal torquing model, but that can be different if one takes non-Gaussian statistics of the tidal fields at late times and on small scales into account. Comparing intrinsic spectra for elliptical galaxies to higher-order effects in gravitational lensing, it is certainly the case that intrinsic alignments are the dominating contributions to shape and size correlations on all scales, whereas other effects such as Born-corrections, lens-lens-coupling or reduced shear corrections are present only at predominantly small scales \citep{krause_weak_2010}. In our model, we use a nonlinear fit to the CDM-spectrum $P(k)$ to include nonlinear scales, which is directly applicable to nonlinear tidal gravitational fields because of the linearity of the Poisson-equation, i.e. the alignment mechanism is still linear, nevertheless the ambient density field generating the tidal gravitational field, is in a stage of nonlinear structure formation.}

Now, the decomposition with Pauli-matrices makes it possible to write down all ellipticity spectra as contractions of the the possible spectra of the source terms, for lensing,
\begin{equation}
C^{\gamma\gamma}_{AB}(\ell) = \frac{1}{4}\sigma^{(1)}_{ab}\sigma^{(1)}_{cd}C^{\psi_A\psi_B}_{abcd}(\ell) = \frac{\ell^4}{4}C^{\psi_A\psi_B}(\ell),
\end{equation}
for intrinsic alignments,
\begin{equation}
C^{\epsilon\epsilon}_{AB}(\ell) = \frac{1}{4}\sigma^{(1)}_{ab}\sigma^{(1)}_{cd}C^{\varphi_A\varphi_B}_{abcd}(\ell) = \frac{\ell^4}{4}C^{\varphi_A\varphi_B}(\ell),
\end{equation}
and for the cross-correlation between the two,
\begin{equation}
C^{\epsilon\gamma}_{AB}(\ell) = \frac{1}{4}\sigma^{(1)}_{ab}\sigma^{(1)}_{cd}C^{\varphi_A\psi_B}_{abcd}(\ell) = \frac{\ell^4}{4}C^{\varphi_A\psi_B}(\ell).
\end{equation}
A measurement of the shape correlations is limited by a Poissonian shape noise contribution,
\begin{equation}
N_{AB}^\mathrm{shape}(\ell) = \sigma^2_\mathrm{shape}\frac{n_\mathrm{tomo}}{\bar{n}}\delta_{AB},
\end{equation}
with a value of $\sigma_\mathrm{shape} = 0.4$ and the number density $\bar{n} = 4.727\times 10^8$ galaxies per steradian typical for Euclid-studies. It is straightforward to show that of the 20 possible spectra 10 are in fact nonzero, and that certain consistency relations hold, for instance $\bra\kappa\kappa^\prime\ket = \bra\gamma_+\gamma_+^\prime\ket + \bra\gamma_\times\gamma_\times^\prime\ket$ as well as $\bra ss^\prime\ket = \bra\epsilon_+\epsilon_+^\prime\ket + \bra\epsilon_\times\epsilon_\times^\prime\ket$, in any coordinate frame.

\begin{figure}
\centering
\includegraphics[scale=0.45]{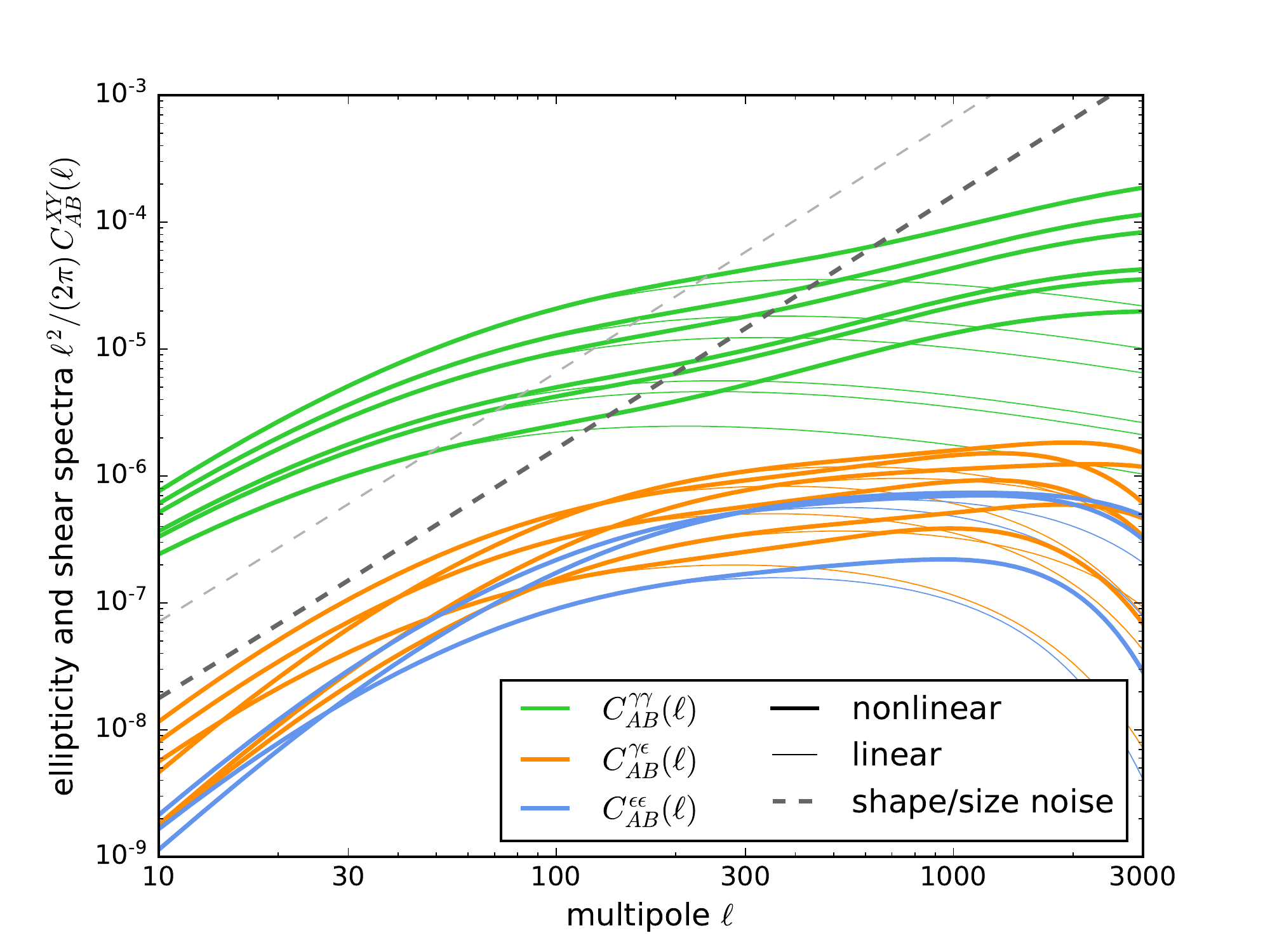}
\caption{Shape-shape correlations as a function of multipole order $\ell$, separated by gravitational lensing $C_{AB}^{\gamma\gamma}(\ell)$, intrinsic size correlations $C_{AB}^{\epsilon\epsilon}(\ell)$ and the cross-correlation $C_{AB}^{\gamma\epsilon}(\ell)$ (of which we show the absolute value), with the Poissonian noise contributions $N_{AB}^\mathrm{shape}(\ell)$ (dark grey) and $N_{AB}^\mathrm{size}(\ell)$ (light grey, a factor of 4 higher) in comparison, for Euclid's redshift distribution and tomography with 3 bins, for a $\Lambda$CDM-cosmology with an alignment parameter $D_{IA}=-10^{-5}~(\mathrm{Mpc}/h)^2$ on a mass scale $M = 10^{12}M_\odot/h$, corresponding to a virial velocity of $\sigma\simeq10^5\mathrm{m}/\mathrm{s}$. Thick and thin lines indicate a nonlinear and linear spectrum, respectively.}
\label{fig:shapeshape}
\end{figure}

The resulting extrinsic and intrinsic shape spectra are shown for a tomographic survey in Fig.~\ref{fig:shapeshape}: Intrinsic shape correlations are relevant at intermediate multipoles, but are surpassed by one to two orders of magnitude by weak lensing-induced shape correlations, for realistic values of the alignment parameter $D_{IA}$. Intrinsic and extrinsic shapes are anti-correlated, and the cross-correlation is modulating the spectra over much wider multipole ranges. In fulfilment of the Cauchy-Schwarz-inequality, the cross-correlation has values between the pure lensing and intrinsic alignment effect. \spirou{The alignment parameter $D_{IA}$ was chosen to be $10^{-5}~(\mathrm{Mpc}/h)^2$, and scales proportional to $\sigma^2$, where $\sigma=10^5\mathrm{m}/\mathrm{s}$ would be a typical value for a Milky Way-sized object with $10^{12} M_\odot/h$: Increasing the velocity dispersion (where $\sigma\propto M^{1/3}$ due to the viral law) requires a larger alignment parameter $D_{IA}$. This value of the alignment parameter is chosen lower than the value measured by \citet{tugendhat_angular_2018} in CFHTLenS-data, with a value of $10^{-4}~(\mathrm{Mpc}/h)^2$ for $D_{IA}$, but with an uncertainty to which mass scale this parameter actually corresponds, making direct comparisons difficult: There are, in fact, many relevant issues to consider: First of all, the choice of a mass scale determines the smoothing scale on which the CDM-spectrum $P(k)$ for computing tidal fields is cut off, because tidal shear field fluctuations on scales smaller than the galaxy itself can not be relevant to the alignment process. Secondly, the mass scale enters the alignment parameter through the arguments discussed in the introduction, and because mass functions are commonly strongly decreasing with mass, the choice of the lower mass limit matters tremendously when considering averaged values of virial quantities and consequently, of the alignment parameter. Lastly, there is an implicit mass dependence generated by the strong dependence of the induced ellipticity change with the S{\'e}rsic-index, as more massive galaxies tend to have higher S{\'e}rsic-indices, i.e. $n\simeq 4$ for ordinary ellipticals versus $n\simeq 1$ for dwarf ellipticals.}

Compared to the IllustrisTNG-simulation \citep{Zjupa_tng_2020}, where the alignment parameter as a constant of proportionality is measured directly in the relation between ellipticity and tidal shear, our value for $D_{IA}$ is higher by a factor of 4, because the measurement of the ambient tidal shear field contains a contribution from the local matter density and disregards biasing effects. Currently, there are still large uncertainties concerning the value and its dependence on galaxy mass as well as a possible evolution in redshift and galaxy biasing, such that we decided to use an intermediate value. A direct measurement of shape alignments in the IllustrisTNG simulation without a differentiation between galaxy types carried out by \citet{hilbert_intrinsic_2017} yields a higher value of $D_{IA}\simeq1.5\times10^{-4}~(\mathrm{Mpc}/h)^2$, but a direct comparison is difficult as galaxy biasing plays certainly a role in correlations as a function of physical separation but less so in line of sight-averaged quantities. Given these arguments we settle for a conservative choice of $10^{-5}~(\mathrm{Mpc}/h)^2$ for $D_{IA}$ and discuss the implications of different parameter values in Sect.~\ref{sect_fisher}. \spirou{On the other side, modelling of intrinsic alignments with effective field theories \cite{PhysRevD.100.103506,fang_fast-pt_2017} would be, in contrast to our analytical approach, able to simulate the combined effect of alignment and biasing, at the expense of a potentially larger number of alignment parameters.}

The shape correlations on very small scales would be dominated by spiral galaxies, for Euclid's redshift distribution and with the assumption of the tidal torquing model this would be the case on multipoles above $\ell\simeq 300$. As in this model the ellipticities are proportional to the quadratic tidal shear field one would not expect for Gaussian fields a cross correlation with the shapes of elliptical galaxies nor with lensing, making the shapes of spiral galaxies statistically uncorrelated.

In a similar manner as in the previous section, one obtains the size spectra from contracting the possible spectra of the source terms, for lensing,
\begin{equation}
C^{\kappa\kappa}_{AB}(\ell) = \frac{1}{4}\sigma^{(0)}_{ab}\sigma^{(0)}_{cd}C^{\psi_A\psi_B}_{abcd}(\ell) = \frac{\ell^4}{4}C^{\psi_A\psi_B}(\ell),
\end{equation}
for intrinsic alignments,
\begin{equation}
C^{ss}_{AB}(\ell) = \frac{1}{4}\sigma^{(0)}_{ab}\sigma^{(0)}_{cd}C^{\varphi_A\varphi_B}_{abcd}(\ell) = \frac{\ell^4}{4}C^{\varphi_A\varphi_B}(\ell),
\end{equation}
and again, for the cross-correlation between the two,
\begin{equation}
C^{s\kappa}_{AB}(\ell) = \frac{1}{4}\sigma^{(0)}_{ab}\sigma^{(0)}_{cd}C^{\varphi_A\psi_B}_{abcd}(\ell) = \frac{\ell^4}{4}C^{\varphi_A\psi_B}(\ell),
\end{equation}
i.e. all size-spectra are equal to their shape-counterparts. In the estimation process, there is a constant, diagonal noise contribution
\begin{equation}
N_{AB}^\mathrm{size}(\ell) = \sigma^2_\mathrm{size} \frac{n_\mathrm{tomo}}{\bar{n}}\delta_{AB},
\end{equation}
with the size noise $\sigma_\mathrm{size} = 0.8$.

Fig.~\ref{fig:shapeshape} shows at the same time the intrinsic and extrinsic \BG{spectra of galaxy shapes}, as they would result from a tomographic survey \spirou{similar to the Euclid lensing survey}. In fact, as a consequence of the linear alignment model and the linearity of weak lensing the size-correlations are identical to the shape-correlations, including the anti-correlation between intrinsic and extrinsic size: \spirou{The $GI$-term arising when correlating lensing convergence with intrinsic size is negative due to exactly the same reasons as lensing shear is anti-correlated with intrinsic ellipticity.} 

\begin{figure}
\centering
\includegraphics[scale=0.45]{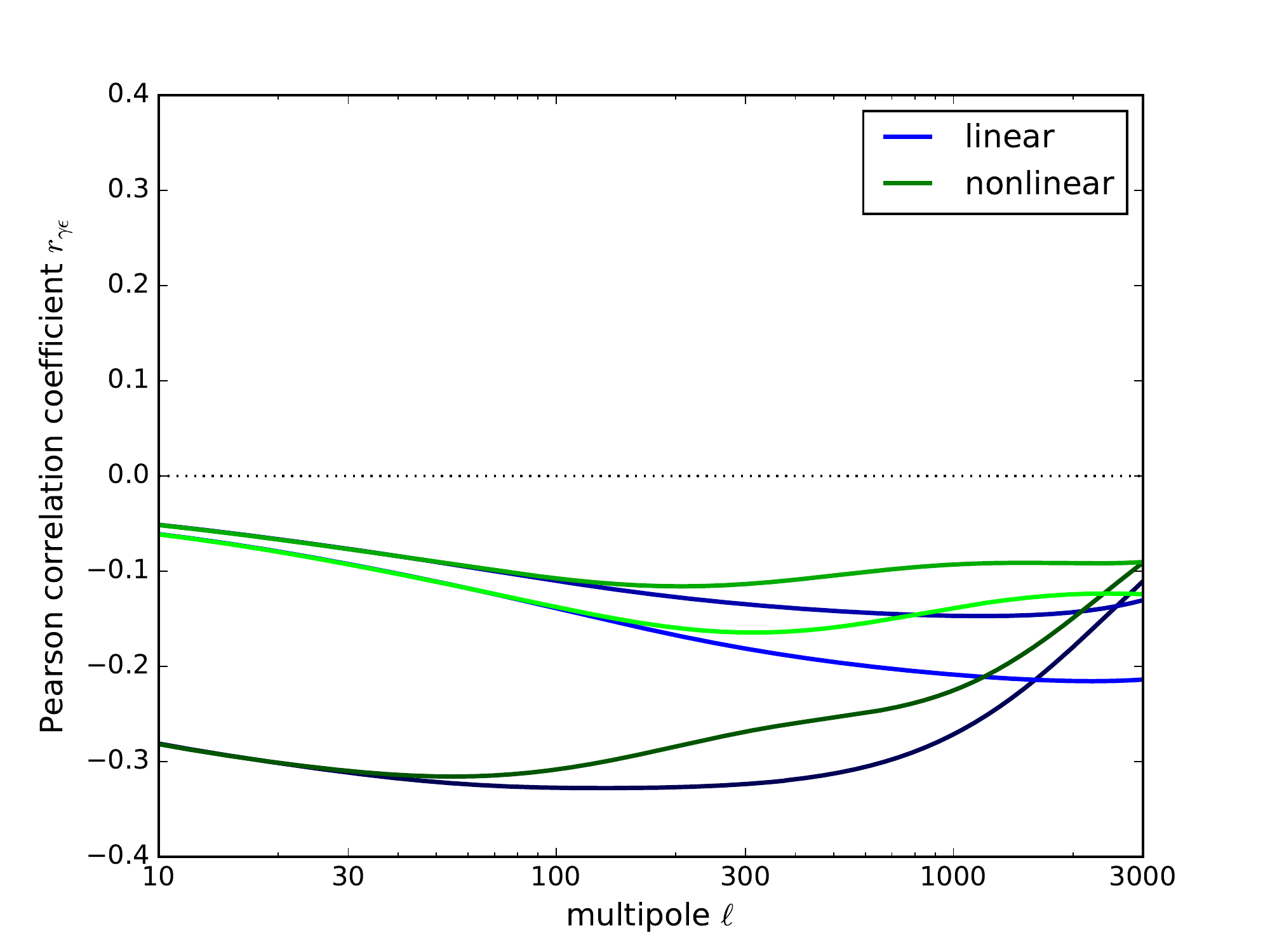}
\caption{Pearson correlation coefficients $r_{\gamma\epsilon}(\ell)$ as a function of multipole order $\ell$. The curves correspond to low redshift (bottom pair), intermediate redshift (top pair) and high redshift (centre pair), and contrast linear (blue lines) and nonlinear (green lines) spectra.}
\label{fig:pearson}
\end{figure}

Given the fact that there is a slightly higher uncertainty in the measurement of angular size in comparison to shape one can already now expect that the corresponding signal to noise-ratios for size-correlations are slightly inferior to shapes. These statements rely on the fact that the same alignment parameter $D_{IA}$ is relevant for both shapes and sizes, as the linear alignment model would suggest. Similarly, we show in Fig.~\ref{fig:pearson} the Pearson correlation coefficient $r_{\gamma\epsilon}(\ell)$ as a function of multipole $\ell$,
\begin{equation}
r_{\gamma\epsilon}(\ell) = 
\frac{C^{\gamma\epsilon}_{AA}(\ell)}{\sqrt{C^{\gamma\gamma}_{AA}(\ell)\: C^{\epsilon\epsilon}_{AA}(\ell)}},
\label{eqn_pearson}
\end{equation}
where we would like to emphasise that the Pearson-coefficients for shapes and sizes are identical, $r_{\gamma\epsilon}(\ell) = r_{\kappa s}{\ell}$. \spirou{The Pearson coefficient shows how statistically independent cross-correlations are from auto-correlations. In equation~(\ref{eqn_pearson}), this coefficient $r_{\gamma\epsilon}(\ell)$ is zero if there is no correlation, 1 if the correlation is perfect, and -1 if there is perfect anti-correlation. The boundedness of the values of $r_{\gamma\epsilon}(\ell)$ is a consequence of the Cauchy-Schwarz-inequality, which states that $\left|C^{\gamma\epsilon}_{AA}(\ell)\right|\leq\sqrt{C^{\gamma\gamma}_{AA}C^{\epsilon\epsilon}_{AA}(\ell)}$. Again, the negative values of $r_{\gamma\epsilon}(\ell)$ indicate the negative sign of the $GI$-terms, for shape- and size-correlations alike. We set the bin-indices equal, $A = B$, because only in this case $C^{\epsilon\epsilon}_{AB}(\ell)$ and $C^{ss}_{AB}(\ell)$ are unequal to zero. The values for $r_{\gamma\epsilon}(\ell)$ suggest that there is in fact redundancy in the spectra.} 

Finally, we compute the cross-correlations between galaxy shapes and sizes, for lensing
\begin{equation}
C_{AB}^{\kappa\gamma}(\ell) = \frac{1}{4}\sigma^{(0)}_{ab}\sigma^{(1)}_{cd}C^{\psi_A\psi_B}_{abcd}(\ell) = \frac{\ell^4}{4}C^{\psi_A\psi_B}(\ell)
\end{equation}
for intrinsic alignments,
\begin{equation}
C_{AB}^{s\epsilon}(\ell) = \frac{1}{4}\sigma^{(0)}_{ab}\sigma^{(1)}_{cd}C^{\varphi_A\varphi_B}_{abcd}(\ell) = \frac{\ell^4}{4}C^{\varphi_A\varphi_B}(\ell)
\end{equation}
and for the cross-correlation between lensing and alignments,
\begin{align}
C_{AB}^{\kappa\epsilon}(\ell) & = \frac{1}{4}\sigma^{(0)}_{ab}\sigma^{(1)}_{cd}C^{\psi_A\varphi_B}_{abcd}(\ell) = \frac{\ell^4}{4}C^{\psi_A\varphi_B}(\ell)\\
C_{AB}^{s\gamma}(\ell) & = \frac{1}{4}\sigma^{(0)}_{ab}\sigma^{(1)}_{cd}C^{\psi_A\varphi_B}_{abcd}(\ell) = \frac{\ell^4}{4}C^{\varphi_A\psi_B}(\ell),
\end{align}
where due to the independence of the errors in the shape and size correlations one does not have to deal with a noise contribution when estimating spectra. Effectively, the cross-correlations between shape and size look identical to the autocorrelations, but in their estimation process there is no noise term, if statistical independence of the two measurement processes for shape and size is given. \BG{This can be seen in e.g.~\cite{heavens_combining_2013} and \cite{alsing_weak_2014}, who as well estimate the amount of shape- and size-noise.}

\section{information content of shape and size correlations}\label{sect_fisher}
For quantifying the information content of intrinsic size and shape correlations in comparison to weak lensing convergence and shear we use the Fisher-matrix formalism. Arranging the measurements of galaxy shapes and sizes into a data vector yields the data covariance matrix,
\begin{equation}
C =
\left(
\begin{array}{cc}
C^{\epsilon\epsilon}_{AB}(\ell) + 2C^{\epsilon\gamma}_{AB}(\ell) + C^{\gamma\gamma}_{AB}(\ell) + N^\mathrm{shape}_{AB} & 
C^{s\epsilon}_{AB^\prime}(\ell) + C^{s\gamma}_{AB^\prime}(\ell) + C^{\kappa\epsilon}_{AB^\prime}(\ell) + C^{\kappa\gamma}_{AB^\prime}(\ell) \\
C^{s\epsilon}_{A^\prime B}(\ell) + C^{s\gamma}_{A^\prime B}(\ell) + C^{\kappa\epsilon}_{A^\prime B}(\ell) + C^{\kappa\gamma}_{A^\prime B}(\ell) & 
C^{ss}_{A^\prime B^\prime}(\ell) + 2C^{s\kappa}_{A^\prime B^\prime}(\ell) + C^{\kappa\kappa}_{A^\prime B^\prime}(\ell) + N^\mathrm{size}_{A^\prime B^\prime}
\end{array}
\right)
\end{equation}
Given the similarities between the shape and size correlations allows to rewrite the covariance matrix as
\begin{equation}
C = \left(
\begin{array}{cc}
\frac{\ell^4}{4}\left(C^{\varphi_A\varphi_B}(\ell)+2C^{\varphi_A\psi_B}(\ell)+C^{\psi_A\psi_B}(\ell)\right) + N^\mathrm{shape}_{AB} & 
\frac{\ell^4}{4}\left(C^{\varphi_A\varphi_{B^\prime}}(\ell)+2C^{\varphi_A\psi_{B^\prime}}(\ell)+C^{\psi_A\psi_{B^\prime}}(\ell)\right)\\
\frac{\ell^4}{4}\left(C^{\varphi_{A^\prime}\varphi_B}(\ell)+2C^{\varphi_{A^\prime}\psi_B}(\ell)+C^{\psi_{A^\prime}\psi_B}(\ell)\right) & 
\frac{\ell^4}{4}\left(C^{\varphi_{A^\prime}\varphi_{B^\prime}}(\ell)+2C^{\varphi_{A^\prime}\psi_{B^\prime}}(\ell)+C^{\psi_{A^\prime}\psi_{B^\prime}}(\ell)\right) + N^\mathrm{size}_{A^\prime B^\prime}
\end{array}
\right),
\end{equation}
which is dangerously close to being singular, underlining the degeneracy between the shape- and size measurements: \spirou{In fact, without the noise contributions $N^\mathrm{shape}_{AB}(\ell)$ and $N^\mathrm{size}_{AB}(\ell)$ the covariance matrix would have a vanishing determinant, $\mathrm{det}C$, and would be singular: This would exactly correspond to the cosmic variance limit, in which the two measurements can not be combined in a sensible way.} Already at this stage one should expect that a combined measurement of shear and size does not yield strong improvements of the signal to noise-ratio alone, and given the fact that the same potentials are involved with identical physical dependencies on cosmology, resulting Fisher-matrices will be very similar. We use the Fisher-matrix formalism as a quick way to quantify the fundamental sensitivities and degeneracies, while noting that the non-Gaussian shape of the likelihood matters in most cases and that tools for dealing with non-Gaussian likelihoods analytically exist \citep{takada_impact_2009, sellentin_non-gaussian_2015}.

The Fisher-matrix $F_{\mu\nu}$ for a tomographic survey assumes the generic form
\begin{equation}
F_{\mu\nu} = f_\mathrm{sky}\sum_\ell\frac{2\ell+1}{2}\mathrm{tr}\left(C^{-1}\partial_\mu S\:C^{-1}\partial_\nu S\right)
\end{equation}
where we implicitly assume a full sky coverage by having independent Fourier-modes. \spirou{At this point it is worth emphasising that we choose to work with the Fisher-matrix that quantifies the compatibility of an ensemble of modes $\epsilon_{\ell m}$ and $\gamma_{\ell m}$ of the ellipticity- and shear field, all with vanishing mean, with the variance predicted by the spectra, and for the modes $s_{\ell m}$ and $\kappa_{\ell m}$ in complete analogy.} Similarly, we define the signal to noise ratio $\Sigma$,
\begin{equation}
\Sigma^2 = f_\mathrm{sky}\sum_\ell\frac{2\ell+1}{2}\mathrm{tr}\left(C^{-1}S\:C^{-1}S\right),
\label{eqn_s2n}
\end{equation}
with the noiseless spectrum $S(\ell)$ of which the signal strength is sought. For the case of Euclid, we extend the summation over the multipoles from $\ell=10$ to $\ell=3000$, \spirou{and we are assuming for simplicity a full-sky coverage with no correlations between different multipoles, which would typically arise in the case of incomplete sky coverage. Instead, we scale down the signal with a sky coverage of $\sqrt{f_\mathrm{sky}}$ in a Poissonian manner, and argue that ignoring the correlations between different multipoles is a small error because most of the signal originates on small angular scales, where these correlations are weak.We set the number of tomographic bins to $n_\mathrm{tomo} = 5$ to demonstrate the behaviour and scaling of the alignment signal relative to the lensing signal.}

\begin{figure}
\centering
\includegraphics[scale=0.45]{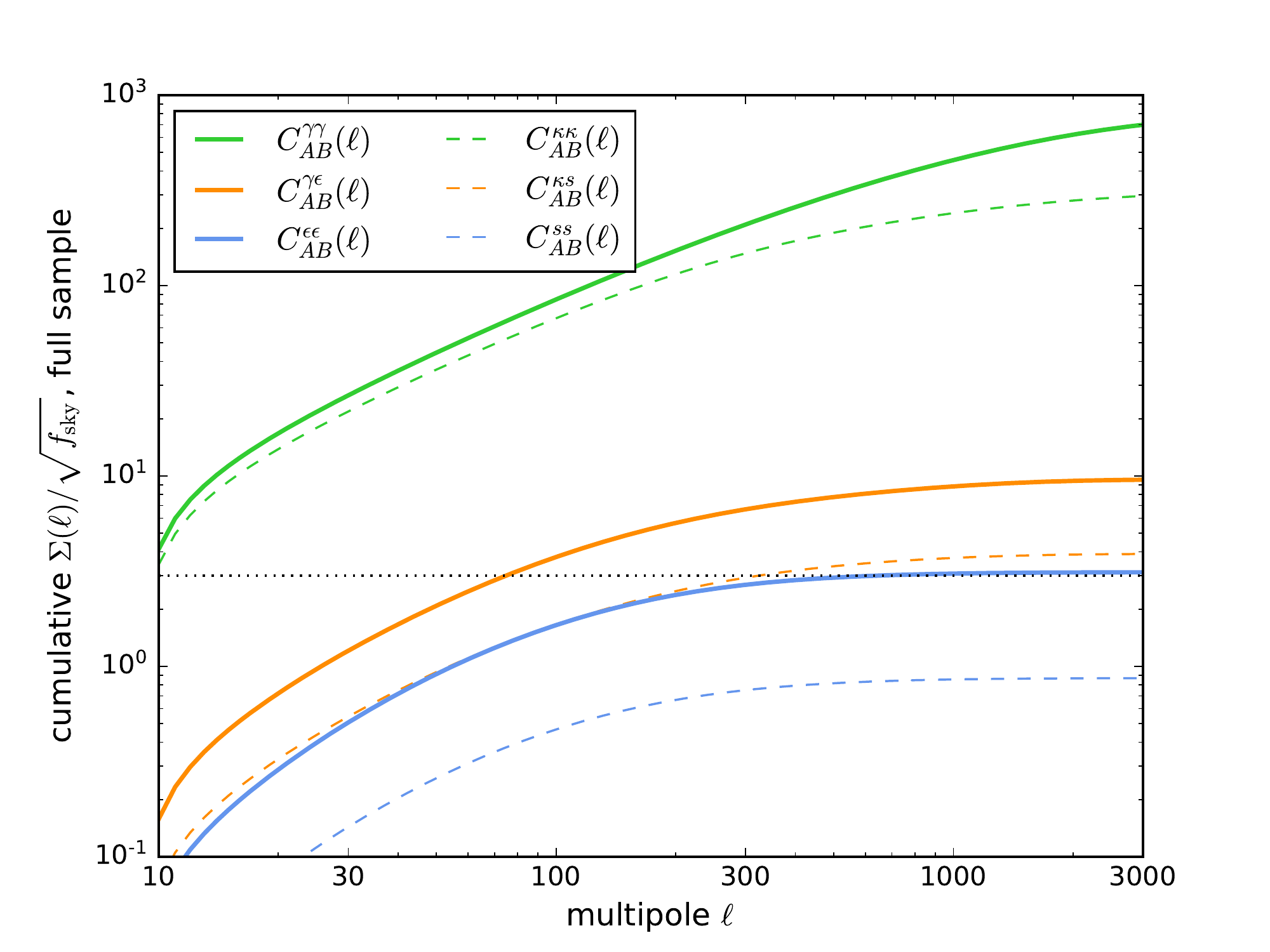}
\caption{Cumulative signal to noise-ratio $\Sigma(\ell)/\sqrt{f_\mathrm{sky}}$ for Euclid 5-bin tomography for measuring shape correlations and intrinsic size correlations, for the full galaxy sample.}
\label{fig:s2n_all}
\end{figure}

Clearly, not all galaxies are ellipticals for which the tidal alignment model would apply, but only a fraction of $q\simeq 1/3$ of them. Therefore, we compute two values for the signal to noise-ratio $\Sigma$: First, we weight the $GI$-type spectra by a factor $q$, and the $II$-type spectra by a factor $q^2$ relative to the $GG$-term, as lensing operates on all galaxies identically irrespective of their type. These numbers for $\Sigma$ would correspond to estimates of the spectra from the full data set and indicate the level of significance by which the shape or size correlations are incompatible with a pure gravitational lensing model. Fig.~\ref{fig:s2n_all} quantifies the signal to noise-ratio $\Sigma$ for measuring intrinsic shape and intrinsic size correlations: We compute the signal to noise-ratio for a measurement of the $II$ and $GI$-terms in both shape- and size correlations in the presence of the full cosmic variance, which is dominated by gravitational lensing, i.e. by the $GG$-terms. As expected, lensing-induced shape correlations are measurable at a higher signal to noise-ratio compared to size correlations, but both are easily within the reach of Euclid. The signal to noise ratio suggests that $GI$-type terms are detectable in shape correlations and perhaps marginally in size correlations, and $II$-terms are marginally detectable, with intrinsic shape correlations being the least disappointing. Because the covariance in equation~(\ref{eqn_s2n}) is by far dominated by weak lensing and by the shape noise and size noise contributions it will be the case that $\Sigma$ is proportional to $\sqrt{D_{IA}}$ for the $GI$-terms and to $D_{IA}$ for the $II$-terms, and other values than $D_{IA}\simeq -10^{-5}~(\mathrm{Mpc}/h)^2$ than the one adopted here will be directly reflected by the signal to noise-prediction. 

As the $II$-terms are proportional to $D_{IA}^2$ and the $GI$-terms proportional to $D_{IA}$, the inverse $\Sigma^{-1}$ of the signal to noise-ratio is at the same time the relative error $D_{IA}/\sigma_{D_{IA}}$ on the alignment parameter $D_{IA}$ for the $GI$-terms, and the absolute error $\sigma = 1/(2\Sigma)$ on $D_{IA}$ for the $II$-terms. This suggests that measurements of the alignment parameter can be carried out at the level of a few ten percent, so the investigation of trends with galaxy mass, type or redshift seem feasible. We have chosen a rather conservative value for $D_{IA}$, nothing precludes the usage of a strategy to boost intrinsic alignments relative to lensing. As for the morphological mix of spiral and elliptical galaxies we conclude that the signal to noise ratios are likewise proportional to $q$ for the $GI$-terms and to $q^2$ for the $II$-terms, such that effectively the combined parameter $q\times D_{IA}$ is determined through a measurement. In the same way as adopting higher values for the alignment parameter $D_{IA}$, a higher fraction of elliptical galaxies $q$ would be reflected in the signal to noise-ratio $\Sigma$.

\begin{figure}
\centering
\includegraphics[scale=0.45]{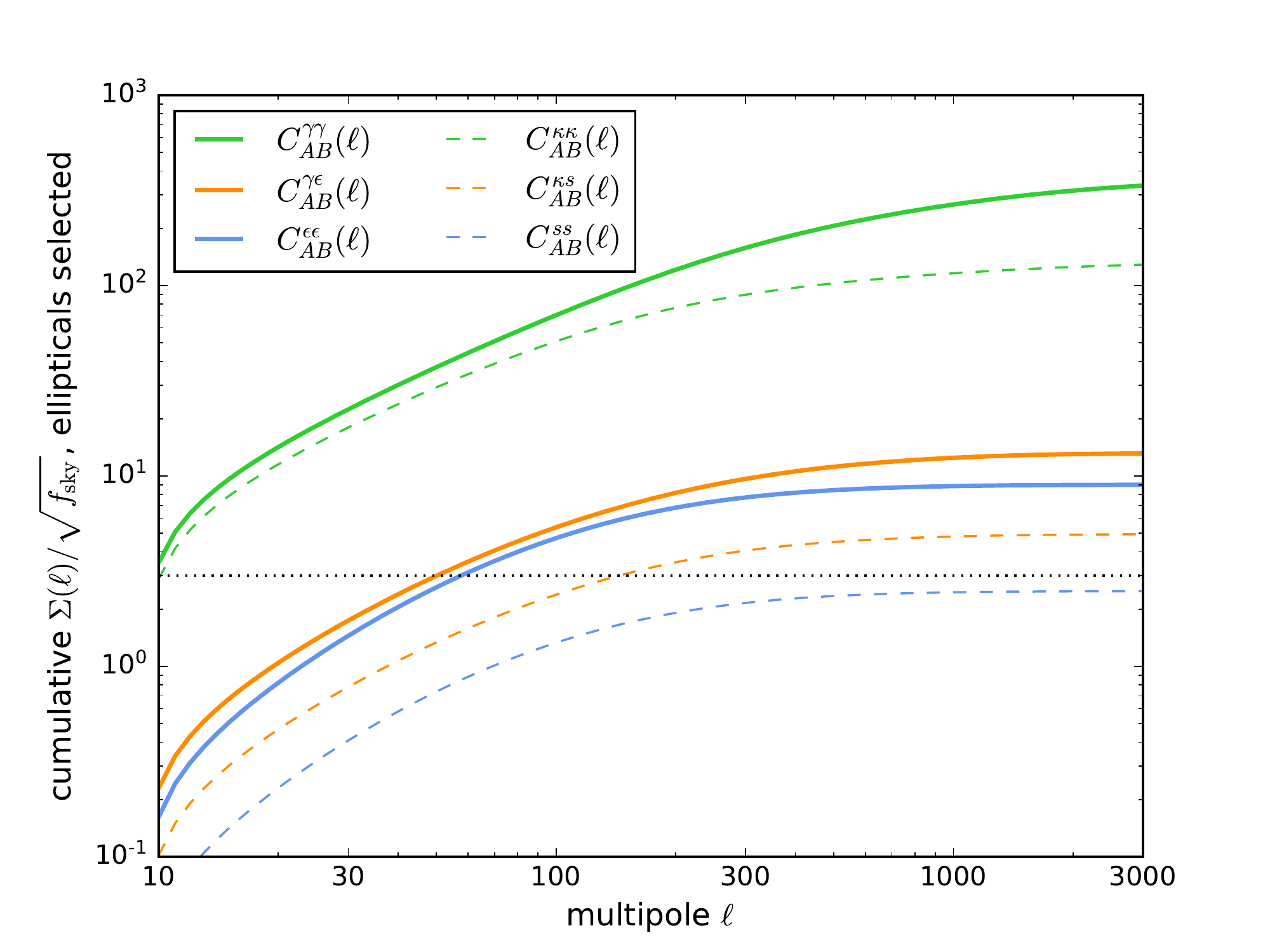}
\caption{Cumulative signal to noise-ratio $\Sigma(\ell)/\sqrt{f_\mathrm{sky}}$ for Euclid 5-bin tomography for measuring shape correlations and size correlations, for a case when elliptical galaxies are selected for the estimation of correlations.}
\label{fig:s2n_elliptical}
\end{figure}

On the other hand one could pursue the strategy to pre-select elliptical galaxies on the basis of their colours or morphologies and to measure the shape- and size correlations on the resulting, reduced data set. In this case, effectively, the total number of galaxies $\bar{n}$ is reduced by $q$ and the number of galaxy pairs by $q^2$, leading to an increased Poissonian noise term, which becomes larger by a factor of $q$. Consequently, the signal strength for weak lensing is much weaker, as it is estimated from a much smaller number of galaxies, but the ratio of the amplitudes between intrinsic alignment and lensing is smaller compared to the previous case: \spirou{In short, one has a cleaner data set because all galaxy pairs carry the intrinsic alignment signal, but this comes at the expense of having lower galaxy numbers and a higher Poisson noise in the estimates for all spectra. For the case of a misidentification of galaxies we refer the reader to \citet{tugendhat_statistical_2018}, where a formalism is presented based on probabilities of misidentification of the first and second kind.} The resulting numbers are shown in Fig.~\ref{fig:s2n_elliptical}, where the overall higher shape and size noise terms decrease the significance, but vice versa, the amplitude of the intrinsic correlations relative to those of lensing are higher, such that a feasible strategy for measuring intrinsic shape correlations could be to measure the $GI$-terms and the $II$-terms with a selected sample of elliptical galaxies. The intrinsic size correlations, however, seem to be out of reach with Euclid, no matter the strategy. The attainable signal to noise ratio depends not only on the alignment parameter $D_{IA}$ but also on the mass-scale on which the spectra are smoothed: The two are not independent and should be related through a virial relationship linking velocity dispersion $\sigma^2$ and mass $M$, $\sigma^2 \propto M^{2/3}$, but choosing a smaller mass scale has the consequence that higher multipoles contribute to the signal an increase $\Sigma(\ell)$. The morphological ratio between spiral and elliptical galaxies impacts in this case only on the total number of galaxies and therefore on the shape and size noise amplitude, as in this case too the cosmic variance is lensing-dominated. 

\spirou{While the second strategy delivers directly the significances of the $GI$- and $II$-terms for a sample of elliptical galaxies, we should be careful in pointing out that the first strategy quantifies the significance of a contribution of elliptical galaxies to the total intrinsic shape correlations, to which spiral galaxies contribute as well, albeit only on higher multipoles \citep{tugendhat_angular_2018}.} If alignments of spiral galaxies follows from the quadratic model \citep{crittenden_spin-induced_2001}, their shapes would be statistically uncorrelated with the shapes of elliptical galaxies and they would not generate a cross-correlation with lensing \citep{tugendhat_statistical_2018}, such that both signal to noise ratios would add in quadrature. 

\begin{figure}
\centering
\includegraphics[scale=0.45]{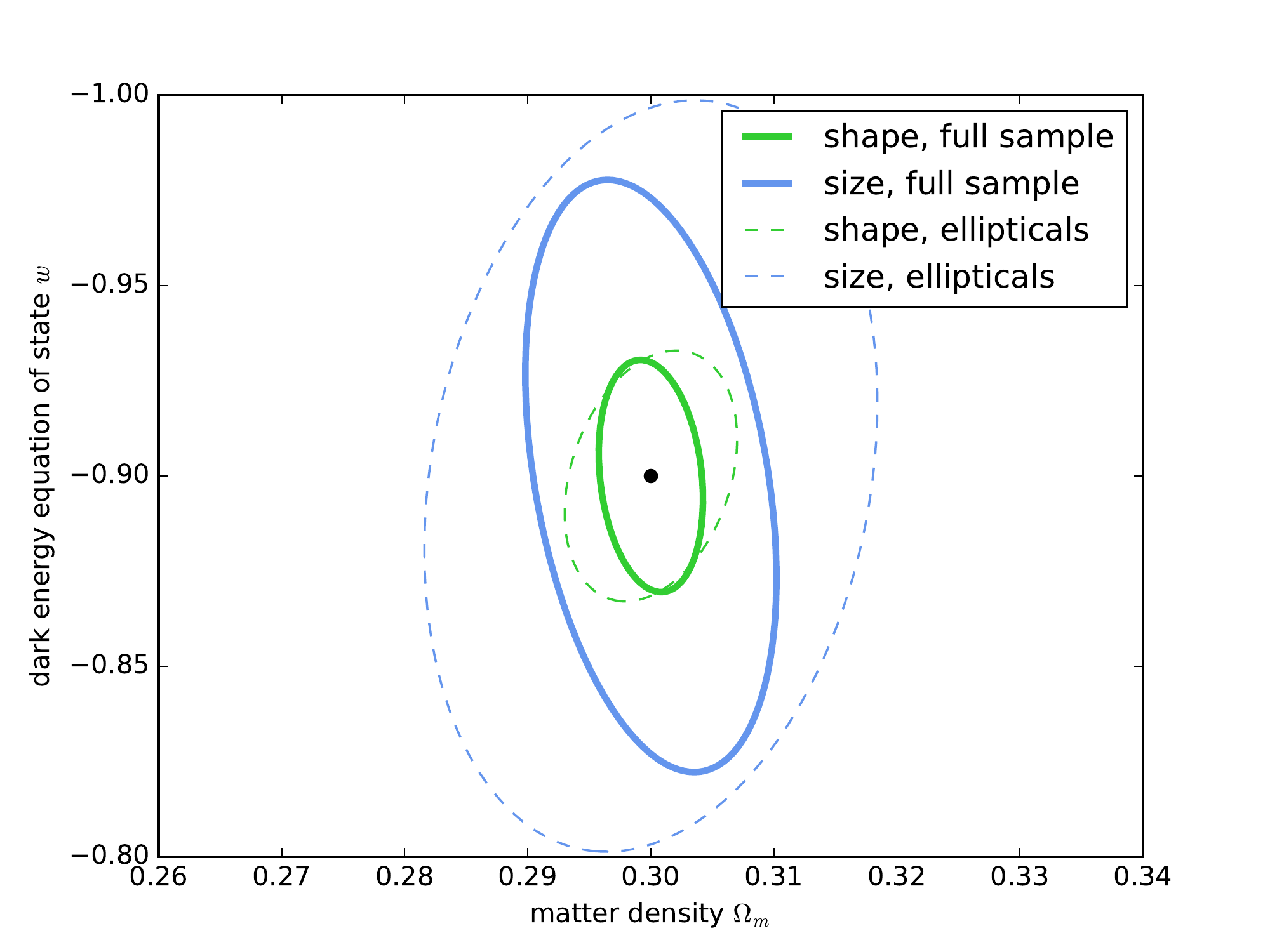}
\caption{Marginalised $1\sigma$-contours from the Fisher-matrix analysis on a standard $w$CDM-cosmology (with $w=-0.9$) for a fixed value $D_{IA}=-10^{-5}~(\mathrm{Mpc}/h)^2$ for the alignment parameter and a smoothing scale of $10^{12}M_\odot/h$: We give contours separated by shape and size correlations, and for the full galaxy sample versus a sample containing only elliptical galaxies.}
\label{fig:fisher}
\end{figure}

\spirou{Fig.~\ref{fig:fisher} shows constraints on a $w$CDM-cosmology from galaxy shapes and galaxy sizes: It is always the case that using the entire galaxy sample leads to tighter constraints than pre-selecting elliptical galaxies, i.e. the lower Poisson-error dominates over the weaker correlations.} As both observables are probing tidal gravitational fields with identical physical dependences there can not be any fundamental difference in the degeneracies, with the only exception that the noise in the size-measurement is typically larger than the one of the shape-measurement, which effectively cuts off high multipoles from contributing to the signal. We emphasise that the two measurements are highly correlated such that one does not gain an advantage from combining the two. We would argue, however, that there is potential to use shape and size-correlations to investigate deviations from the Newtonian form of the Poisson equation e.g. by modified theories of gravity. \spirou{The relative amplitudes of the $GG$-, $GI$- and $II$-terms are sensitive to gravitational slip, because one compares the effect of gravitational potential on relativistic (lensing) to nonrelativistic (stars inside a galaxy) test particles, with the potential advantage over the combination of lensing with peculiar velocities that the measurement combines two probes on the same scale.} For this, one needs a very good understanding of the detailed mechanisms of alignment with possibly nonlinear corrections to the tidal alignment model, as well as the scaling behaviour of the alignment parameter with redshift and galaxy mass \citep{hirata_intrinsic_2007}, and possibly different alignment parameters for subpopulations of elliptical galaxies, as the strong dependence on the S{\'e}rsic-index suggests. \foca{Ultimately, one would need to resort to the concept that on galactic scales Newtonian gravity is prevalent, such that the amount of shape (or size) distortion of an elliptical galaxy can be predicted from simulations of galaxy formation and evolution. Then, relative to the Newtonian prediction, modified gravity effects on larger scales can be parameterised and determined by gravitational lensing. Without detailed input from numerical simulations or a much deeper understanding of galaxy scaling relations the degeneracy between gravitational slip and the alignment parameter $D_{IA}$ can not be broken.}

Fundamental degeneracies in the spectra are present between the alignment parameter $D_{IA}$ and $\sigma_8$, which are perfectly degenerate in the linear regime, but the degeneracy is broken by combining $GG$, $GI$ and $II$-terms in the measurement, as they are proportional to $\sigma_8^2$, $\sigma_8^2D_{IA}$ and $\sigma_8^2D_{IA}^2$, respectively. In a  similar way, the proportionality of the lensing spectrum to $\Omega_m^2$ to first order translates to the $GI$-term, which is proportional to $\Omega_m$. The influence of the particular dark energy model by mapping the redshift distribution of the galaxies onto a distribution in comoving distance, is identical for all correlations. Pursuing the two strategies of either keeping the full galaxy sample and down-weighting $GI$-spectra by $q=1/3$ and $II$-terms by $q^2$ yields smaller errors than pre-selecting elliptical galaxies first, because the smaller Poisson-noise, but the second strategy has a higher relative contribution from intrinsic alignments, which start to matter when deriving constraints, as they provide cosmological information. \BG{In both cases we obtain the result that the dominating constraint on the alignment parameter $D_{IA}$ is derived from the $GI$-term due to its higher amplitude compared to the $II$-term, if all other parameters are fixed.} 

\section{summary}\label{sect_summary}
The subject of our investigation were extrinsic and intrinsic shape and size correlations of elliptical galaxies due to weak gravitational lensing and intrinsic alignments\foca{, using an analytical model, without any simulation}. Our starting point was the description of the stellar density of a virialised system through the Jeans-equation, in which we perturb the gravitational potential with an external tidal  field. Under the condition that this field is reasonably weak and the galaxy compact enough, one can compute the response in shape and size of a galaxy in linear approximation in the tidal field,  controlled by the galaxy's velocity dispersion $\sigma^2$. The susceptibility of a galaxy to tidal distortions is highly dependent on the stellar profile: A toy model using the S{\'e}rsic profile family shows a strong increase in the response from exponential profiles to de Vaucouleurs-profiles.

These are our main findings:
\begin{itemize}
\item{Assuming a weakly perturbed Jeans-equilibrium for elliptical galaxies naturally reproduces a linear response of the shape and the size of a galaxy to external tidal gravitational fields, and suggests that the same alignment parameter is responsible for the change in shape and in size. \foca{Nominally, the velocity dispersion $\sigma$ of the galaxy sets the scale for the gravitational field, which is remarkably similar to the quantity $2\Phi/c^2$ in gravitational lensing. With virial equilibrium one can continue to argue that $\sigma^2$ is proportional to $M/R$ with the mass $M$ and the size $R$, such that the ratio $(R/\sigma)^2$, which controls the strength of the tidal interaction, is in fact constant \citep[compare][]{piras_mass_2018}. A mass-dependence of the intrinsic alignment effect would be introduced by the convolution of the tidal shear spectrum with a filter function corresponding to the galaxy mass $M$, and the strong scaling of the expected alignment effect with S{\'e}rsic-index $n$, which commonly tends to increase with galaxy mass. Galaxy biasing would introduce an additional modulation of the intrinsic alignment effect and should be included in particular when comparing intrinsic alignment spectra with straightforward galaxy clustering; in this sense the intrinsic shapes and sizes become weighted clustering spectra. Again, one finds typically higher biases for more massive galaxies, adding another scaling of the results with mass.}}

\item{Using the standard Poisson-equation, the galaxy sizes provide a direct mapping of the ambient matter density, and  the intrinsic and extrinsic shapes and sizes are consistent with each other, \foca{including the factor of two which appears in the relationship between the angular Laplacian of the lensing potential $\psi$ and the weak lensing convergence $\kappa$ for gravitational lensing and the relation between the angular Laplacian of the alignment potential $\Phi$ (which is simply the projected Newtonian gravitational potential) and the intrinsic size $s$ of a galaxy.} To which extent this can be used to probe deviations from Newtonian gravity is largely unclear and depends on a detailed understanding of the astrophysics of the objects, \foca{and ultimately, the constant of proportionality between intrinsic alignment and tidal field has to be determined from simulations because it does not, unlike gravitational lensing, follow from fundamental physics.} When using shape- and size-correlations as cosmological probes, the Poisson equation causes them to contain only degenerate information, and there is a direct mapping between $GG$, $GI$ and $II$-type terms. In addition, the shape and size-correlations are highly degenerate to the point where size correlations become redundant in comparison to the stronger and more sensitive shape correlations. We note, however, that size correlations can provide an alternative method for mapping out the matter distribution.}

\item{Similar to the case of shape correlations, one obtains a completely diagonal autocorrelation for the intrinsic sizes, $C^{ss}_{AB}(\ell)\propto\delta_{AB}$ and a non-diagonal cross-correlation between size and convergence, $C^{s\kappa}_{AB}(\ell)$. The non-diagonal part of the lensing signal only contains $GG$ and $GI$, but never $II$-terms \citep{jain_cross-correlation_2003, takada_tomography_2004, huterer_nulling_2005}, and in principle nulling- and boosting techniques \citep{joachimi_removal_2009, 2010arXiv1009.2024J, joachimi_intrinsic_2010} are applicable to size-correlations as well.}

\item{Computing a forecast for Euclid we find that intrinsic shape- and size-correlations as well as their cross-correlations with lensing are measurable. Typical signal to noise-ratios obtained for 5-bin tomography are with Euclid range around 10 for $C^{\gamma\epsilon}_{AB}(\ell)$- and $C^{\epsilon\epsilon}_{AB}(\ell)$-correlations, while size correlations are more difficult to detect. Simulating two strategies, measuring correlations in the full galaxy sample or pre-selecting elliptical galaxies first, showed that the latter could be able to make $C^{\epsilon\epsilon}_{AB}(\ell)$-correlations detectable. Our forecasts uses a conservative value for the alignment parameter, $D_{IA}\simeq -10^{-5}~(\mathrm{Mpc}/h)^2$, which should strongly depend on the mass scale \citep{piras_mass_2018} and potentially on the profile shape as well. With this particular value of $D_{IA}$, among the size correlations, only $C^{\kappa s}_{AB}(\ell)$ could yield a marginal detection. But since the intrinsic signal is directly proportional to $D_{IA}$, increasing $D_{IA}$ by a factor 3-4 would change this result \foca{and would shift all intrinsic spectra within reach of Euclid}.}

\item{\spirou{Investigating the dependence of the spectra on the fundamental parameters of the cosmological model with a standard Fisher-matrix analysis shows that intrinsic shape and size-correlations have essentially identical parameter dependencies, irrespective of whether the mechanism is gravitational lensing or intrinsic alignments, similar to the results presented by \citet{alsing_weak_2014}.} Typically, the shape-measurement yields smaller Poissonian errors compared to the size estimation, such that the value of the errors is smaller in a size measurement. A combination of the two does not yield significant improvements due to the large covariance between the two measurements. Nevertheless, since they are complementary, the two measurements can provide a consistency test for General Relativity on cosmological scales. We pursued two strategies, which consist in pre-selecting the elliptical galaxies, which increases the noise due to reducing the data, or keeping the full galaxy sample and down-weighting the $GI$- and $II$-terms with the fraction of elliptical galaxies. The first strategy yields tighter errors, but  the second strategy picks up stronger contributions from the $GI$- and $II$-terms to the Fisher-matrix, which in turn are very similar to galaxy clustering correlations. \spirou{Estimation biases as they would arise in fitting lensing models to data that includes both lensing and intrinsic alignments, should arise in complete analogy in the size-sector as they have been demonstrated to matter for shape correlations, \citep{joachimi_simultaneous_2010, kirk_galaxy_2015, joachimi_galaxy_2015, krause_impact_2016}.}}

\end{itemize}

In the future, we plan to investigate the usability of both types of shape and size spectra for designing specific tests of gravity, for instance for Vainshtein-type screening mechanisms \citep{kirk_optimising_2011, tessore_weak_2015}, which would manifest themselves in differences between the intrinsic and extrinsic shape and size spectra. Likewise, there is the question whether measurements of the velocity dispersion can help to disentangle intrinsic size from lensing shear, as the size effect causes galaxies with the same velocity dispersion to appear systematically larger in underdense regions, and through velocity dispersion a common baseline could be established. \foca{Again, we point out that these studies would need to be informed with a prior on $D_{IA}$ obtained from simulations.} In addition, we point out that the susceptibility $\int\dd r\:r^5\rho(r)$ of a stellar system with density $\rho$ could differ for subclasses of elliptical galaxies giving rise to different effective alignment parameters $D_{IA}$. Let us briefly comment on possible intrinsic-size and shape effects arising at second order: Similar to lens-lens coupling one can expect a $B$-mode generation if lensing shear acts on a correlated intrinsic ellipticity field \citep[similar to][]{cooray_second-order_2002}, and if lensing deflection shifts the galaxies to new positions \citep{giahi_evolution_2013, giahi-saravani_weak_2014}. To what extent spiral galaxies exhibit similar intrinsic size correlations is unclear, and possibly much more dependent on the astrophysics of galaxy formation, beyond models of tidal torquing \citep{schaefer_review:_2009}. Finally, we point out that intrinsic size correlations are straightforward to be implemented in effective field theories of structure formation \citep{fang_fast-pt_2017, vlah_eft_2019}, as they only require the computation of $\Delta\Phi$ on a smoothed field.

\section*{Acknowledgements}
BG thanks the University of Heidelberg for hospitality. BMS likes to thank the Universidad del Valle in Cali, Colombia, for their kind hospitality. We thank Eileen Sophie Giesel and Jolanta Zjupa for spotting mistakes in an early version of the draft. We are also grateful to the anonymous referee for their valuable insights and comments for our paper. BG and RD acknowledge support from the Swiss National Science Foundation.

\section*{Data Availability}
There are no new data associated with this article.
\bibliographystyle{mnras}
\bibliography{references}

\begin{thebibliography}{}
\makeatletter
\relax
\def\mn@urlcharsother{\let\do\@makeother \do\$\do\&\do\#\do\^\do\_\do\%\do\~}
\def\mn@doi{\begingroup\mn@urlcharsother \@ifnextchar [ {\mn@doi@}
  {\mn@doi@[]}}
\def\mn@doi@[#1]#2{\def\@tempa{#1}\ifx\@tempa\@empty \href
  {http://dx.doi.org/#2} {doi:#2}\else \href {http://dx.doi.org/#2} {#1}\fi
  \endgroup}
\def\mn@eprint#1#2{\mn@eprint@#1:#2::\@nil}
\def\mn@eprint@arXiv#1{\href {http://arxiv.org/abs/#1} {{\tt arXiv:#1}}}
\def\mn@eprint@dblp#1{\href {http://dblp.uni-trier.de/rec/bibtex/#1.xml}
  {dblp:#1}}
\def\mn@eprint@#1:#2:#3:#4\@nil{\def\@tempa {#1}\def\@tempb {#2}\def\@tempc
  {#3}\ifx \@tempc \@empty \let \@tempc \@tempb \let \@tempb \@tempa \fi \ifx
  \@tempb \@empty \def\@tempb {arXiv}\fi \@ifundefined
  {mn@eprint@\@tempb}{\@tempb:\@tempc}{\expandafter \expandafter \csname
  mn@eprint@\@tempb\endcsname \expandafter{\@tempc}}}

\bibitem[\protect\citeauthoryear{Abbott et~al.}{Abbott
  et~al.}{2018}]{Abbott:2017wau}
Abbott T. M.~C.,  et~al., 2018, \mn@doi [Phys. Rev. D]
  {10.1103/PhysRevD.98.043526}, 98, 043526

\bibitem[\protect\citeauthoryear{Alsing, Kirk, Heavens  \& Jaffe}{Alsing
  et~al.}{2015}]{alsing_weak_2014}
Alsing J.,  Kirk D.,  Heavens A.,   Jaffe A.,  2015, MNRAS, 452, 1202

\bibitem[\protect\citeauthoryear{Altay, Colberg  \& Croft}{Altay
  et~al.}{2006}]{altay_influence_2006}
Altay G.,  Colberg J.~M.,   Croft R. A.~C.,  2006, \mn@doi [MNRAS]
  {10.1111/j.1365-2966.2006.10555.x}, 370, 1422

\bibitem[\protect\citeauthoryear{Amara \& Refregier}{Amara \&
  Refregier}{2007}]{amara_optimal_2007}
Amara A.,  Refregier A.,  2007, \mn@doi [MNRAS]
  {10.1111/j.1365-2966.2007.12271.x}, 381, 1018

\bibitem[\protect\citeauthoryear{Amendola et~al.}{Amendola
  et~al.}{2018}]{Amendola:2016saw}
Amendola L.,  et~al., 2018, Living Reviews in Relativity, 21, 345

\bibitem[\protect\citeauthoryear{Bailin \& Steinmetz}{Bailin \&
  Steinmetz}{2005}]{bailin_internal_2005}
Bailin J.,  Steinmetz M.,  2005, \mn@doi [ApJ] {10.1086/430397}, 627, 647

\bibitem[\protect\citeauthoryear{Bartelmann}{Bartelmann}{2010}]{bartelmann_gravitational_2010}
Bartelmann M.,  2010, Classical and Quantum Gravity, 27, 233001

\bibitem[\protect\citeauthoryear{Bartelmann \& Schneider}{Bartelmann \&
  Schneider}{2001}]{bartelmann_weak_2001}
Bartelmann M.,  Schneider P.,  2001, Physics Reports, 340, 291

\bibitem[\protect\citeauthoryear{Bate, Chisari, Codis, Martin, Dubois,
  Devriendt, Pichon  \& Slyz}{Bate et~al.}{2019}]{bate_when_2019}
Bate J.,  Chisari N.~E.,  Codis S.,  Martin G.,  Dubois Y.,  Devriendt J.,
  Pichon C.,   Slyz A.,  2019, \mn@doi [MNRAS] {10.1093/mnras/stz3166}, 491,
  4057

\bibitem[\protect\citeauthoryear{Bernstein}{Bernstein}{2009}]{bernstein_comprehensive_2009}
Bernstein G.~M.,  2009, \mn@doi [ApJ] {10.1088/0004-637X/695/1/652}, 695, 652

\bibitem[\protect\citeauthoryear{Bernstein \& Jain}{Bernstein \&
  Jain}{2004}]{bernstein_dark_2004}
Bernstein G.,  Jain B.,  2004, \mn@doi [ApJ] {10.1086/379768}, 600, 17

\bibitem[\protect\citeauthoryear{Bernstein \& Jarvis}{Bernstein \&
  Jarvis}{2002}]{bernstein_shapes_2002}
Bernstein G.~M.,  Jarvis M.,  2002, \mn@doi [AJ] {10.1086/338085}, 123, 583

\bibitem[\protect\citeauthoryear{Blazek, Mandelbaum, Seljak  \&
  Nakajima}{Blazek et~al.}{2012}]{blazek_separating_2012}
Blazek J.,  Mandelbaum R.,  Seljak U.,   Nakajima R.,  2012, JCAP, 2012, 041

\bibitem[\protect\citeauthoryear{Blazek, MacCrann, Troxel  \& Fang}{Blazek
  et~al.}{2019}]{PhysRevD.100.103506}
Blazek J.~A.,  MacCrann N.,  Troxel M.~A.,   Fang X.,  2019, \mn@doi [Phys.
  Rev. D] {10.1103/PhysRevD.100.103506}, 100, 103506

\bibitem[\protect\citeauthoryear{Brown, Taylor, Hambly  \& Dye}{Brown
  et~al.}{2002}]{brown_measurement_2002}
Brown M.~L.,  Taylor A.~N.,  Hambly N.~C.,   Dye S.,  2002, \mn@doi [MNRAS]
  {10.1046/j.1365-8711.2002.05354.x}, 333, 501

\bibitem[\protect\citeauthoryear{Camelio \& Lombardi}{Camelio \&
  Lombardi}{2015}]{camelio_origin_2015}
Camelio G.,  Lombardi M.,  2015, A+A, 575, A113

\bibitem[\protect\citeauthoryear{Capranico, Merkel  \& Sch{\"a}fer}{Capranico
  et~al.}{2013}]{capranico_intrinsic_2013}
Capranico F.,  Merkel P.~M.,   Sch{\"a}fer B.~M.,  2013, \mn@doi [MNRAS]
  {10.1093/mnras/stt1269}, 435, 194

\bibitem[\protect\citeauthoryear{Casarini, La~Vacca, Amendola, Bonometto  \&
  Macci{\`o}}{Casarini et~al.}{2011}]{casarini_non-linear_2011}
Casarini L.,  La~Vacca G.,  Amendola L.,  Bonometto S.~A.,   Macci{\`o} A.~V.,
  2011, \mn@doi [JCAP] {10.1088/1475-7516/2011/03/026}, 3, 26

\bibitem[\protect\citeauthoryear{Catelan, Kamionkowski  \& Blandford}{Catelan
  et~al.}{2001}]{catelan_intrinsic_2001}
Catelan P.,  Kamionkowski M.,   Blandford R.~D.,  2001, MNRAS, 320, L7

\bibitem[\protect\citeauthoryear{Chang et~al.,}{Chang
  et~al.}{2018}]{chang_dark_2017}
Chang C.,  et~al., 2018, MNRAS, 475, 3165

\bibitem[\protect\citeauthoryear{Chisari \& Dvorkin}{Chisari \&
  Dvorkin}{2013}]{chisari_cosmological_2013}
Chisari N.~E.,  Dvorkin C.,  2013, \mn@doi [JCAP]
  {10.1088/1475-7516/2013/12/029}, 12, 29

\bibitem[\protect\citeauthoryear{Chisari, Mandelbaum, Strauss, Huff  \&
  Bahcall}{Chisari et~al.}{2014a}]{chisari_intrinsic_2014}
Chisari N.~E.,  Mandelbaum R.,  Strauss M.~A.,  Huff E.~M.,   Bahcall N.~A.,
  2014a, \mn@doi [MNRAS] {10.1093/mnras/stu1786}, 445, 726

\bibitem[\protect\citeauthoryear{Chisari et~al.,}{Chisari
  et~al.}{2014b}]{chisari_intrinsic_2015}
Chisari N.~E.,  et~al., 2014b, MNRAS, 454, 2736

\bibitem[\protect\citeauthoryear{Chisari, Dunkley, Miller  \& Allison}{Chisari
  et~al.}{2015}]{chisari_contamination_2015}
Chisari N.~E.,  Dunkley J.,  Miller L.,   Allison R.,  2015, \mn@doi [MNRAS]
  {10.1093/mnras/stv1655}, 453, 682

\bibitem[\protect\citeauthoryear{Chisari et~al.,}{Chisari
  et~al.}{2016}]{chisari_redshift_2016}
Chisari N.~E.,  et~al., 2016, \mn@doi [MNRAS] {10.1093/mnras/stw1409}, 461,
  2702

\bibitem[\protect\citeauthoryear{Cooray \& Hu}{Cooray \&
  Hu}{2001}]{cooray_power_2001}
Cooray A.,  Hu W.,  2001, \mn@doi [ApJ] {10.1086/321376}, 554, 56

\bibitem[\protect\citeauthoryear{Cooray \& Hu}{Cooray \&
  Hu}{2002}]{cooray_second-order_2002}
Cooray A.,  Hu W.,  2002, \mn@doi [ApJ] {10.1086/340892}, 574, 19

\bibitem[\protect\citeauthoryear{Crittenden, Natarajan, Pen  \&
  Theuns}{Crittenden et~al.}{2001}]{crittenden_spin-induced_2001}
Crittenden R.~G.,  Natarajan P.,  Pen U.-L.,   Theuns T.,  2001, \mn@doi [ApJ]
  {10.1086/322370}, 559, 552

\bibitem[\protect\citeauthoryear{Debattista, van~den Bosch, Roskar, Quinn,
  Moore  \& Cole}{Debattista et~al.}{2015}]{debattista_internal_2015}
Debattista V.~P.,  van~den Bosch F.~C.,  Roskar R.,  Quinn T.,  Moore B.,
  Cole D.~R.,  2015, MNRAS, 452, 4094

\bibitem[\protect\citeauthoryear{Douspis, Salvati  \& Aghanim}{Douspis
  et~al.}{2018}]{Douspis:2018xlj}
Douspis M.,  Salvati L.,   Aghanim N.,  2018, \mn@doi [PoS]
  {10.22323/1.335.0037}, EDSU2018, 037

\bibitem[\protect\citeauthoryear{Dubinski}{Dubinski}{1992}]{dubinski_cosmological_1992}
Dubinski J.,  1992, \mn@doi [ApJ] {10.1086/172076}, 401, 441

\bibitem[\protect\citeauthoryear{Fan}{Fan}{2007}]{fan_intrinsic_2007}
Fan Z.-H.,  2007, \mn@doi [ApJ] {10.1086/521182}, 669, 10

\bibitem[\protect\citeauthoryear{Fang, Blazek, {McEwen}  \& Hirata}{Fang
  et~al.}{2017}]{fang_fast-pt_2017}
Fang X.,  Blazek J.~A.,  {McEwen} J.~E.,   Hirata C.~M.,  2017, \mn@doi [JCAP]
  {10.1088/1475-7516/2017/02/030}, pp 030--030

\bibitem[\protect\citeauthoryear{Forero-Romero, Contreras  \&
  Padilla}{Forero-Romero et~al.}{2014}]{forero-romero_cosmic_2014}
Forero-Romero J.~E.,  Contreras S.,   Padilla N.,  2014, \mn@doi [MNRAS]
  {10.1093/mnras/stu1150}, 443, 1090

\bibitem[\protect\citeauthoryear{Ghosh, Durrer  \& Sellentin}{Ghosh
  et~al.}{2018}]{Ghosh:2018nsm}
Ghosh B.,  Durrer R.,   Sellentin E.,  2018, \mn@doi [JCAP]
  {10.1088/1475-7516/2018/06/008}, 1806, 008

\bibitem[\protect\citeauthoryear{Giahi-Saravani \& Sch{\"a}fer}{Giahi-Saravani
  \& Sch{\"a}fer}{2013}]{giahi_evolution_2013}
Giahi-Saravani A.,  Sch{\"a}fer B.~M.,  2013, \mn@doi [MNRAS]
  {10.1093/mnras/sts110}, 428, 1312

\bibitem[\protect\citeauthoryear{Giahi-Saravani \& Sch{\"a}fer}{Giahi-Saravani
  \& Sch{\"a}fer}{2014}]{giahi-saravani_weak_2014}
Giahi-Saravani A.,  Sch{\"a}fer B.~M.,  2014, \mn@doi [MNRAS]
  {10.1093/mnras/stt2016}, 437, 1847

\bibitem[\protect\citeauthoryear{Graham \& Driver}{Graham \&
  Driver}{2005}]{graham_concise_2005}
Graham A.~W.,  Driver S.~P.,  2005, \mn@doi [Publications of the Astronomical
  Society of Australia] {10.1071/AS05001}, 22, 118

\bibitem[\protect\citeauthoryear{Grassi \& Sch{\"a}fer}{Grassi \&
  Sch{\"a}fer}{2014}]{grassi_detecting_2014}
Grassi A.,  Sch{\"a}fer B.~M.,  2014, \mn@doi [MNRAS] {10.1093/mnras/stt2075},
  437, 2632

\bibitem[\protect\citeauthoryear{Hall \& Taylor}{Hall \&
  Taylor}{2014}]{hall_intrinsic_2014}
Hall A.,  Taylor A.,  2014, \mn@doi [MNRAS] {10.1093/mnrasl/slu094}, 443, L119

\bibitem[\protect\citeauthoryear{Heavens}{Heavens}{2003}]{heavens_3d_2003}
Heavens A.,  2003, \mn@doi [MNRAS] {10.1046/j.1365-8711.2003.06780.x}, 343,
  1327

\bibitem[\protect\citeauthoryear{Heavens, Refregier  \& Heymans}{Heavens
  et~al.}{2000}]{heavens_intrinsic_2000}
Heavens A.,  Refregier A.,   Heymans C.,  2000, MNRAS, 319, 649

\bibitem[\protect\citeauthoryear{Heavens, Kitching  \& Taylor}{Heavens
  et~al.}{2006}]{heavens_measuring_2006}
Heavens A.~F.,  Kitching T.~D.,   Taylor A.~N.,  2006, MNRAS, 373, 105

\bibitem[\protect\citeauthoryear{Heavens, Alsing  \& Jaffe}{Heavens
  et~al.}{2013}]{heavens_combining_2013}
Heavens A.,  Alsing J.,   Jaffe A.,  2013, \mn@doi [MNRAS]
  {10.1093/mnrasl/slt045}, 433, L6

\bibitem[\protect\citeauthoryear{Heymans \& Heavens}{Heymans \&
  Heavens}{2003}]{heymans_weak_2003}
Heymans C.,  Heavens A.,  2003, \mn@doi [MNRAS]
  {10.1046/j.1365-8711.2003.06213.x}, 339, 711

\bibitem[\protect\citeauthoryear{Heymans, Brown, Heavens, Meisenheimer, Taylor
  \& Wolf}{Heymans et~al.}{2004}]{heymans_weak_2004}
Heymans C.,  Brown M.,  Heavens A.,  Meisenheimer K.,  Taylor A.,   Wolf C.,
  2004, \mn@doi [MNRAS] {10.1111/j.1365-2966.2004.07264.x}, 347, 895

\bibitem[\protect\citeauthoryear{Heymans et~al.,}{Heymans
  et~al.}{2013}]{heymans_cfhtlens_2013}
Heymans C.,  et~al., 2013, \mn@doi [MNRAS] {10.1093/mnras/stt601}, 432, 2433

\bibitem[\protect\citeauthoryear{Hilbert, Xu, Schneider, Springel, Vogelsberger
   \& Hernquist}{Hilbert et~al.}{2017}]{hilbert_intrinsic_2017}
Hilbert S.,  Xu D.,  Schneider P.,  Springel V.,  Vogelsberger M.,   Hernquist
  L.,  2017, \mn@doi [MNRAS] {10.1093/mnras/stx482}, 468, 790

\bibitem[\protect\citeauthoryear{Hirata \& Seljak}{Hirata \&
  Seljak}{2010}]{hirata_intrinsic_2010}
Hirata C.~M.,  Seljak U.,  2010, \mn@doi [PRD] {10.1103/PhysRevD.82.049901},
  82, 049901

\bibitem[\protect\citeauthoryear{Hirata, Padmanabhan, Seljak, Schlegel  \&
  Brinkmann}{Hirata et~al.}{2004a}]{hirata_cross-correlation_2004}
Hirata C.~M.,  Padmanabhan N.,  Seljak U.,  Schlegel D.,   Brinkmann J.,
  2004a, \mn@doi [PRD] {10.1103/PhysRevD.70.103501}, 70, 103501

\bibitem[\protect\citeauthoryear{Hirata et~al.,}{Hirata
  et~al.}{2004b}]{hirata_galaxy-galaxy_2004}
Hirata C.~M.,  et~al., 2004b, \mn@doi [MNRAS]
  {10.1111/j.1365-2966.2004.08090.x}, 353, 529

\bibitem[\protect\citeauthoryear{Hirata, Mandelbaum, Ishak, Seljak, Nichol,
  Pimbblet, Ross  \& Wake}{Hirata et~al.}{2007}]{hirata_intrinsic_2007}
Hirata C.~M.,  Mandelbaum R.,  Ishak M.,  Seljak U.,  Nichol R.,  Pimbblet
  K.~A.,  Ross N.~P.,   Wake D.,  2007, \mn@doi [MNRAS]
  {10.1111/j.1365-2966.2007.12312.x}, 381, 1197

\bibitem[\protect\citeauthoryear{Hu}{Hu}{2001}]{hu_dark_2001}
Hu W.,  2001, \mn@doi [PRD] {10.1103/PhysRevD.65.023003}, 65, 023003

\bibitem[\protect\citeauthoryear{Hu}{Hu}{2002}]{hu_dark_2002}
Hu W.,  2002, \mn@doi [PRD] {10.1103/PhysRevD.66.083515}, 66, 083515

\bibitem[\protect\citeauthoryear{Hu \& Tegmark}{Hu \&
  Tegmark}{1999}]{hu_weak_1999}
Hu W.,  Tegmark M.,  1999, \mn@doi [ApJL] {10.1086/311947}, 514, L65

\bibitem[\protect\citeauthoryear{Huff \& Graves}{Huff \&
  Graves}{2014}]{Huff:2011cq}
Huff E.~M.,  Graves G.~J.,  2014, \mn@doi [Astrophys. J. Lett.]
  {10.1088/2041-8205/780/2/L16}, 780, L16

\bibitem[\protect\citeauthoryear{{Hui} \& {Zhang}}{{Hui} \&
  {Zhang}}{2002}]{2002astro.ph..5512H}
{Hui} L.,  {Zhang} J.,  2002, arXiv e-prints, \href
  {https://ui.adsabs.harvard.edu/abs/2002astro.ph..5512H} {pp
  astro--ph/0205512}

\bibitem[\protect\citeauthoryear{Huterer}{Huterer}{2002}]{huterer_weak_2002}
Huterer D.,  2002, \mn@doi [PRD] {10.1103/PhysRevD.65.063001}, 65, 063001

\bibitem[\protect\citeauthoryear{Huterer}{Huterer}{2010}]{huterer_weak_2010}
Huterer D.,  2010, \mn@doi [Gen. Relat. Grav.] {10.1007/s10714-010-1051-z}, 42,
  2177

\bibitem[\protect\citeauthoryear{Huterer \& Takada}{Huterer \&
  Takada}{2005}]{huterer_calibrating_2005}
Huterer D.,  Takada M.,  2005, \mn@doi [Astroparticle Physics]
  {10.1016/j.astropartphys.2005.02.006}, 23, 369

\bibitem[\protect\citeauthoryear{Huterer \& White}{Huterer \&
  White}{2005}]{huterer_nulling_2005}
Huterer D.,  White M.,  2005, \mn@doi [PRD] {10.1103/PhysRevD.72.043002}, 72,
  043002

\bibitem[\protect\citeauthoryear{Jain \& Seljak}{Jain \&
  Seljak}{1997}]{jain_cosmological_1997}
Jain B.,  Seljak U.,  1997, ApJ, 484, 560

\bibitem[\protect\citeauthoryear{Jain \& Taylor}{Jain \&
  Taylor}{2003}]{jain_cross-correlation_2003}
Jain B.,  Taylor A.,  2003, \mn@doi [PRL] {10.1103/PhysRevLett.91.141302}, 91,
  141302

\bibitem[\protect\citeauthoryear{Jee, Tyson, Schneider, Wittman, Schmidt  \&
  Hilbert}{Jee et~al.}{2013}]{jee_cosmic_2013}
Jee M.~J.,  Tyson J.~A.,  Schneider M.~D.,  Wittman D.,  Schmidt S.,   Hilbert
  S.,  2013, \mn@doi [ApJ] {10.1088/0004-637X/765/1/74}, 765, 74

\bibitem[\protect\citeauthoryear{Jing}{Jing}{2002}]{jing_intrinsic_2002}
Jing Y.~P.,  2002, \mn@doi [MNRAS] {10.1046/j.1365-8711.2002.05899.x}, 335, L89

\bibitem[\protect\citeauthoryear{Joachimi \& Bridle}{Joachimi \&
  Bridle}{2010}]{joachimi_simultaneous_2010}
Joachimi B.,  Bridle S.~L.,  2010, \mn@doi [A+A] {10.1051/0004-6361/200913657},
  523, A1

\bibitem[\protect\citeauthoryear{Joachimi \& Schneider}{Joachimi \&
  Schneider}{2009}]{joachimi_removal_2009}
Joachimi B.,  Schneider P.,  2009, \mn@doi [A+A] {10.1051/0004-6361/200912420},
  507, 105

\bibitem[\protect\citeauthoryear{{Joachimi} \& {Schneider}}{{Joachimi} \&
  {Schneider}}{2010a}]{2010arXiv1009.2024J}
{Joachimi} B.,  {Schneider} P.,  2010a, arXiv e-prints, \href
  {https://ui.adsabs.harvard.edu/abs/2010arXiv1009.2024J} {p. arXiv:1009.2024}

\bibitem[\protect\citeauthoryear{Joachimi \& Schneider}{Joachimi \&
  Schneider}{2010b}]{joachimi_intrinsic_2010}
Joachimi B.,  Schneider P.,  2010b, \mn@doi [A+A]
  {10.1051/0004-6361/201014482}, 517, A4

\bibitem[\protect\citeauthoryear{Joachimi, Mandelbaum, Abdalla  \&
  Bridle}{Joachimi et~al.}{2011}]{joachimi_constraints_2011}
Joachimi B.,  Mandelbaum R.,  Abdalla F.~B.,   Bridle S.~L.,  2011, \mn@doi
  [A+A] {10.1051/0004-6361/201015621}, 527, A26

\bibitem[\protect\citeauthoryear{Joachimi et~al.,}{Joachimi
  et~al.}{2015}]{joachimi_galaxy_2015}
Joachimi B.,  et~al., 2015, \mn@doi [Space Science Reviews]
  {10.1007/s11214-015-0177-4}, 193, 1

\bibitem[\protect\citeauthoryear{Johnston et~al.,}{Johnston
  et~al.}{2018}]{johnston_kids+gama:_2018}
Johnston H.,  et~al., 2018, A+A, 624, A30

\bibitem[\protect\citeauthoryear{Joudaki et~al.,}{Joudaki
  et~al.}{2017}]{joudaki_cfhtlens_2017}
Joudaki S.,  et~al., 2017, \mn@doi [MNRAS] {10.1093/mnras/stw2665}, 465, 2033

\bibitem[\protect\citeauthoryear{Joudaki et~al.}{Joudaki
  et~al.}{2018}]{Joudaki:2017zdt}
Joudaki S.,  et~al., 2018, \mn@doi [Mon. Not. Roy. Astron. Soc.]
  {10.1093/mnras/stx2820}, 474, 4894

\bibitem[\protect\citeauthoryear{Joudaki et~al.}{Joudaki
  et~al.}{2020}]{Joudaki:2019pmv}
Joudaki S.,  et~al., 2020, \mn@doi [Astron. Astrophys.]
  {10.1051/0004-6361/201936154}, 638, L1

\bibitem[\protect\citeauthoryear{Kaiser}{Kaiser}{1992}]{kaiser_weak_1992}
Kaiser N.,  1992, \mn@doi [ApJ] {10.1086/171151}, 388, 272

\bibitem[\protect\citeauthoryear{{Kayo} \& {Takada}}{{Kayo} \&
  {Takada}}{2013}]{2013arXiv1306.4684K}
{Kayo} I.,  {Takada} M.,  2013, arXiv e-prints, \href
  {https://ui.adsabs.harvard.edu/abs/2013arXiv1306.4684K} {p. arXiv:1306.4684}

\bibitem[\protect\citeauthoryear{Kayo, Takada  \& Jain}{Kayo
  et~al.}{2013}]{kayo_information_2013}
Kayo I.,  Takada M.,   Jain B.,  2013, \mn@doi [MNRAS] {10.1093/mnras/sts340},
  429, 344

\bibitem[\protect\citeauthoryear{Kiessling et~al.,}{Kiessling
  et~al.}{2015}]{kiessling_galaxy_2015}
Kiessling A.,  et~al., 2015, \mn@doi [Space Science Reviews]
  {10.1007/s11214-015-0203-6}, 193, 67

\bibitem[\protect\citeauthoryear{Kilbinger}{Kilbinger}{2015}]{kilbinger_cosmology_2015}
Kilbinger M.,  2015, \mn@doi [Reports on Progress in Physics]
  {10.1088/0034-4885/78/8/086901}, 78, 086901

\bibitem[\protect\citeauthoryear{Kilbinger et~al.,}{Kilbinger
  et~al.}{2009}]{kilbinger_dark-energy_2009}
Kilbinger M.,  et~al., 2009, \mn@doi [A+A] {10.1051/0004-6361/200811247}, 497,
  677

\bibitem[\protect\citeauthoryear{Kilbinger et~al.,}{Kilbinger
  et~al.}{2013}]{kilbinger_cfhtlens:_2013}
Kilbinger M.,  et~al., 2013, \mn@doi [MNRAS] {10.1093/mnras/stt041}, 430, 2200

\bibitem[\protect\citeauthoryear{Kirk, Bridle  \& Schneider}{Kirk
  et~al.}{2010}]{kirk_impact_2010}
Kirk D.,  Bridle S.,   Schneider M.,  2010, \mn@doi [MNRAS]
  {10.1111/j.1365-2966.2010.17213.x}, 408, 1502

\bibitem[\protect\citeauthoryear{Kirk, Laszlo, Bridle  \& Bean}{Kirk
  et~al.}{2011}]{kirk_optimising_2011}
Kirk D.,  Laszlo I.,  Bridle S.,   Bean R.,  2011, MNRAS, 430, 197

\bibitem[\protect\citeauthoryear{Kirk et~al.,}{Kirk
  et~al.}{2015a}]{kirk_galaxy_2015}
Kirk D.,  et~al., 2015a, \mn@doi [Space Science Reviews]
  {10.1007/s11214-015-0213-4}, 193, 139

\bibitem[\protect\citeauthoryear{Kirk et~al.,}{Kirk
  et~al.}{2015b}]{kirk_cross-correlation_2015}
Kirk D.,  et~al., 2015b, MNRAS, 459, 21

\bibitem[\protect\citeauthoryear{Kitching, Alsing, Heavens, Jimenez, {McEwen}
  \& Verde}{Kitching et~al.}{2017}]{kitching_limits_2016}
Kitching T.~D.,  Alsing J.,  Heavens A.~F.,  Jimenez R.,  {McEwen} J.~D.,
  Verde L.,  2017, MNRAS, 469, 2737

\bibitem[\protect\citeauthoryear{Krause \& Hirata}{Krause \&
  Hirata}{2010}]{krause_weak_2010}
Krause E.,  Hirata C.~M.,  2010, \mn@doi [A+A] {10.1051/0004-6361/200913524},
  523, A28

\bibitem[\protect\citeauthoryear{Krause, Eifler  \& Blazek}{Krause
  et~al.}{2016}]{krause_impact_2016}
Krause E.,  Eifler T.,   Blazek J.,  2016, \mn@doi [MNRAS]
  {10.1093/mnras/stv2615}, 456, 207

\bibitem[\protect\citeauthoryear{{LSST Dark Energy Science
  Collaboration}}{{LSST Dark Energy Science
  Collaboration}}{2012}]{Abate:2012za}
{LSST Dark Energy Science Collaboration} 2012, preprint, \href
  {http://adsabs.harvard.edu/abs/2012arXiv1211.0310L} {} (\mn@eprint {arXiv}
  {1211.0310})

\bibitem[\protect\citeauthoryear{Lange et~al.,}{Lange
  et~al.}{2015}]{10.1093/mnras/stu2467}
Lange R.,  et~al., 2015, \mn@doi [Monthly Notices of the Royal Astronomical
  Society] {10.1093/mnras/stu2467}, 447, 2603

\bibitem[\protect\citeauthoryear{Larsen \& Challinor}{Larsen \&
  Challinor}{2016}]{larsen_intrinsic_2016}
Larsen P.,  Challinor A.,  2016, \mn@doi [MNRAS] {10.1093/mnras/stw1645}, 461,
  4343

\bibitem[\protect\citeauthoryear{Lee \& Erdogdu}{Lee \&
  Erdogdu}{2007}]{lee_alignments_2007}
Lee J.,  Erdogdu P.,  2007, ApJ, 671, 1248

\bibitem[\protect\citeauthoryear{Lee \& Pen}{Lee \&
  Pen}{2008}]{lee_nonlinear_2007}
Lee J.,  Pen U.-L.,  2008, ApJ, 681, 798

\bibitem[\protect\citeauthoryear{{MacCrann}, Zuntz, Bridle, Jain  \&
  Becker}{{MacCrann} et~al.}{2014}]{maccrann_cosmic_2014}
{MacCrann} N.,  Zuntz J.,  Bridle S.,  Jain B.,   Becker M.~R.,  2014, MNRAS,
  451, 2877

\bibitem[\protect\citeauthoryear{Mackey, White  \& Kamionkowski}{Mackey
  et~al.}{2002}]{mackey_theoretical_2002}
Mackey J.,  White M.,   Kamionkowski M.,  2002, \mn@doi [MNRAS]
  {10.1046/j.1365-8711.2002.05337.x}, 332, 788

\bibitem[\protect\citeauthoryear{Mandelbaum et~al.,}{Mandelbaum
  et~al.}{2011}]{mandelbaum_wigglez_2011}
Mandelbaum R.,  et~al., 2011, \mn@doi [MNRAS]
  {10.1111/j.1365-2966.2010.17485.x}, 410, 844

\bibitem[\protect\citeauthoryear{Massey et~al.,}{Massey
  et~al.}{2013}]{massey_origins_2013}
Massey R.,  et~al., 2013, \mn@doi [MNRAS] {10.1093/mnras/sts371}, 429, 661

\bibitem[\protect\citeauthoryear{Mellier}{Mellier}{1999}]{mellier_probing_1999}
Mellier Y.,  1999, \mn@doi [ARAA] {10.1146/annurev.astro.37.1.127}, 37, 127

\bibitem[\protect\citeauthoryear{Merkel \& Schaefer}{Merkel \&
  Schaefer}{2013}]{merkel_intrinsic_2013}
Merkel P.~M.,  Schaefer B.~M.,  2013, \mn@doi [MNRAS] {10.1093/mnras/stt1151},
  434, 1808

\bibitem[\protect\citeauthoryear{Merkel \& Schaefer}{Merkel \&
  Schaefer}{2017}]{merkel_imitating_2017}
Merkel P.~M.,  Schaefer B.~M.,  2017, \mn@doi [MNRAS] {10.1093/mnras/stx1664},
  471, 2431

\bibitem[\protect\citeauthoryear{{Mortonson}, {Weinberg}  \&
  {White}}{{Mortonson} et~al.}{2013}]{2014arXiv1401.0046M}
{Mortonson} M.~J.,  {Weinberg} D.~H.,   {White} M.,  2013, arXiv e-prints,
  \href {https://ui.adsabs.harvard.edu/abs/2014arXiv1401.0046M} {p.
  arXiv:1401.0046}

\bibitem[\protect\citeauthoryear{Munshi, Valageas, van Waerbeke  \&
  Heavens}{Munshi et~al.}{2008}]{munshi_cosmology_2008}
Munshi D.,  Valageas P.,  van Waerbeke L.,   Heavens A.,  2008, \mn@doi
  [Physics Reports] {10.1016/j.physrep.2008.02.003}, 462, 67

\bibitem[\protect\citeauthoryear{Munshi, Coles  \& Kilbinger}{Munshi
  et~al.}{2014}]{munshi_tomography_2014}
Munshi D.,  Coles P.,   Kilbinger M.,  2014, \mn@doi [JCAP]
  {10.1088/1475-7516/2014/04/004}, 04, 004

\bibitem[\protect\citeauthoryear{Pahwa et~al.,}{Pahwa
  et~al.}{2016}]{pahwa_alignment_2016}
Pahwa I.,  et~al., 2016, \mn@doi [MNRAS] {10.1093/mnras/stv2930}, 457, 695

\bibitem[\protect\citeauthoryear{Pandya et~al.,}{Pandya
  et~al.}{2019}]{pandya_can_2019}
Pandya V.,  et~al., 2019, \mn@doi [MNRAS] {10.1093/mnras/stz2129}, 488, 5580

\bibitem[\protect\citeauthoryear{Peacock \& Heavens}{Peacock \&
  Heavens}{1985}]{peacock_statistics_1985}
Peacock J.~A.,  Heavens A.~F.,  1985, MNRAS, 217, 805

\bibitem[\protect\citeauthoryear{Pedersen, Yao, Ishak  \& Zhang}{Pedersen
  et~al.}{2020}]{Pedersen:2019wfp}
Pedersen E.~M.,  Yao J.,  Ishak M.,   Zhang P.,  2020, \mn@doi [Astrophys. J.
  Lett.] {10.3847/2041-8213/aba51b}, 899, L5

\bibitem[\protect\citeauthoryear{Piras, Joachimi, Sch{\"a}fer, Bonamigo,
  Hilbert  \& van Uitert}{Piras et~al.}{2018}]{piras_mass_2018}
Piras D.,  Joachimi B.,  Sch{\"a}fer B.~M.,  Bonamigo M.,  Hilbert S.,   van
  Uitert E.,  2018, \mn@doi [MNRAS] {10.1093/mnras/stx2846}, 474, 1165

\bibitem[\protect\citeauthoryear{Reischke \& Sch{\"a}fer}{Reischke \&
  Sch{\"a}fer}{2019}]{reischke_environmental_2018}
Reischke R.,  Sch{\"a}fer B.~M.,  2019, JCAP, 04, 031

\bibitem[\protect\citeauthoryear{Schaefer}{Schaefer}{2009}]{schaefer_review:_2009}
Schaefer B.~M.,  2009, \mn@doi [IJMPD] {10.1142/S0218271809014388}, 18, 173

\bibitem[\protect\citeauthoryear{Sch{\"a}fer \& Merkel}{Sch{\"a}fer \&
  Merkel}{2012}]{schafer_galactic_2012}
Sch{\"a}fer B.~M.,  Merkel P.~M.,  2012, \mn@doi [MNRAS]
  {10.1111/j.1365-2966.2011.20224.x}, 421, 2751

\bibitem[\protect\citeauthoryear{Schmitz, Hirata, Blazek  \& Krause}{Schmitz
  et~al.}{2018}]{schmitz_time_2018}
Schmitz D.~M.,  Hirata C.~M.,  Blazek J.,   Krause E.,  2018, JCAP, 07, 030

\bibitem[\protect\citeauthoryear{Schneider \& Bridle}{Schneider \&
  Bridle}{2010}]{schneider_halo_2010}
Schneider M.~D.,  Bridle S.,  2010, \mn@doi [MNRAS]
  {10.1111/j.1365-2966.2009.15956.x}, 402, 2127

\bibitem[\protect\citeauthoryear{Schneider et~al.,}{Schneider
  et~al.}{2013}]{schneider_galaxy_2013}
Schneider M.~D.,  et~al., 2013, \mn@doi [MNRAS] {10.1093/mnras/stt855}, 433,
  2727

\bibitem[\protect\citeauthoryear{Sellentin \& Sch{\"a}fer}{Sellentin \&
  Sch{\"a}fer}{2015}]{sellentin_non-gaussian_2015}
Sellentin E.,  Sch{\"a}fer B.~M.,  2015, MNRAS, 456, 1645

\bibitem[\protect\citeauthoryear{Semboloni, Hoekstra, Schaye, van Daalen  \&
  {McCarthy}}{Semboloni et~al.}{2011}]{semboloni_quantifying_2011}
Semboloni E.,  Hoekstra H.,  Schaye J.,  van Daalen M.~P.,   {McCarthy} I.~G.,
  2011, \mn@doi [MNRAS] {10.1111/j.1365-2966.2011.19385.x}, 417, 2020

\bibitem[\protect\citeauthoryear{S{\'e}rsic}{S{\'e}rsic}{1963}]{sersic_influence_1963}
S{\'e}rsic J.~L.,  1963, Boletin de la Asociacion Argentina de Astronomia La
  Plata Argentina, 6, 41

\bibitem[\protect\citeauthoryear{Singh, Mandelbaum  \& More}{Singh
  et~al.}{2015}]{Singh:2014kla}
Singh S.,  Mandelbaum R.,   More S.,  2015, \mn@doi [Mon. Not. Roy. Astron.
  Soc.] {10.1093/mnras/stv778}, 450, 2195

\bibitem[\protect\citeauthoryear{Takada \& Jain}{Takada \&
  Jain}{2009}]{takada_impact_2009}
Takada M.,  Jain B.,  2009, \mn@doi [MNRAS] {10.1111/j.1365-2966.2009.14504.x},
  395, 2065

\bibitem[\protect\citeauthoryear{Takada \& White}{Takada \&
  White}{2004}]{takada_tomography_2004}
Takada M.,  White M.,  2004, \mn@doi [ApJL] {10.1086/381870}, 601, L1

\bibitem[\protect\citeauthoryear{Takahashi, Oguri, Sato  \& Hamana}{Takahashi
  et~al.}{2011}]{takahashi_probability_2011}
Takahashi R.,  Oguri M.,  Sato M.,   Hamana T.,  2011, \mn@doi [ApJ]
  {10.1088/0004-637X/742/1/15}, 742, 15

\bibitem[\protect\citeauthoryear{{Taruya} \& {Okumura}}{{Taruya} \&
  {Okumura}}{2020}]{2020ApJ...891L..42T}
{Taruya} A.,  {Okumura} T.,  2020, \mn@doi [\apjl] {10.3847/2041-8213/ab7934},
  \href {https://ui.adsabs.harvard.edu/abs/2020ApJ...891L..42T} {891, L42}

\bibitem[\protect\citeauthoryear{Tenneti, Mandelbaum, Di~Matteo, Feng  \&
  Khandai}{Tenneti et~al.}{2014}]{tenneti_galaxy_2014}
Tenneti A.,  Mandelbaum R.,  Di~Matteo T.,  Feng Y.,   Khandai N.,  2014,
  \mn@doi [MNRAS] {10.1093/mnras/stu586}, 441, 470

\bibitem[\protect\citeauthoryear{Tenneti, Singh, Mandelbaum, Matteo, Feng  \&
  Khandai}{Tenneti et~al.}{2015}]{tenneti_intrinsic_2015}
Tenneti A.,  Singh S.,  Mandelbaum R.,  Matteo T.~D.,  Feng Y.,   Khandai N.,
  2015, \mn@doi [MNRAS] {10.1093/mnras/stv272}, 448, 3522

\bibitem[\protect\citeauthoryear{Tessore, Winther, Metcalf, Ferreira  \&
  Giocoli}{Tessore et~al.}{2015}]{tessore_weak_2015}
Tessore N.,  Winther H.~A.,  Metcalf R.~B.,  Ferreira P.~G.,   Giocoli C.,
  2015, \mn@doi [JCAP] {10.1088/1475-7516/2015/10/036}, pp 036--036

\bibitem[\protect\citeauthoryear{Thomas, Bruni  \& Wands}{Thomas
  et~al.}{2015}]{thomas_relativistic_2014}
Thomas D.~B.,  Bruni M.,   Wands D.,  2015, JCAP, 09, 021

\bibitem[\protect\citeauthoryear{Troxel \& Ishak}{Troxel \&
  Ishak}{2012}]{troxel_self-calibration_2012}
Troxel M.~A.,  Ishak M.,  2012, \mn@doi [MNRAS]
  {10.1111/j.1365-2966.2012.20987.x}, 423, 1663

\bibitem[\protect\citeauthoryear{Troxel \& Ishak}{Troxel \&
  Ishak}{2015}]{troxel_intrinsic_2015}
Troxel M.~A.,  Ishak M.,  2015, \mn@doi [Physics Reports]
  {10.1016/j.physrep.2014.11.001}, 558, 1

\bibitem[\protect\citeauthoryear{Tugendhat \& Schaefer}{Tugendhat \&
  Schaefer}{2018}]{tugendhat_angular_2018}
Tugendhat T.~M.,  Schaefer B.~M.,  2018, \mn@doi [MNRAS]
  {10.1093/mnras/sty323}, 476, 3460

\bibitem[\protect\citeauthoryear{Tugendhat, Reischke  \& Schaefer}{Tugendhat
  et~al.}{2020}]{tugendhat_statistical_2018}
Tugendhat T.~M.,  Reischke R.,   Schaefer B.~M.,  2020, \mn@doi [MNRAS]
  {10.1093/mnras/staa641}, 494, 2969

\bibitem[\protect\citeauthoryear{Vlah, Chisari  \& Schmidt}{Vlah
  et~al.}{2020}]{vlah_eft_2019}
Vlah Z.,  Chisari N.~E.,   Schmidt F.,  2020, JCAP, 01, 025

\bibitem[\protect\citeauthoryear{White}{White}{2004}]{white_baryons_2004}
White M.,  2004, \mn@doi [Astroparticle Physics]
  {10.1016/j.astropartphys.2004.06.001}, 22, 211

\bibitem[\protect\citeauthoryear{Yao, Ishak, Lin  \& Troxel}{Yao
  et~al.}{2017}]{yao_effects_2017}
Yao J.,  Ishak M.,  Lin W.,   Troxel M.~A.,  2017, JCAP, 10, 056

\bibitem[\protect\citeauthoryear{Yao, Pedersen, Ishak, Zhang, Agashe, Xu  \&
  Shan}{Yao et~al.}{2019a}]{yao_separating_2019}
Yao J.,  Pedersen E.~M.,  Ishak M.,  Zhang P.,  Agashe A.,  Xu H.,   Shan H.,
  2019a, arXiv 1911.01582

\bibitem[\protect\citeauthoryear{Yao, Ishak  \& Troxel}{Yao
  et~al.}{2019b}]{yao_self-calibration_2018}
Yao J.,  Ishak M.,   Troxel M.~A.,  2019b, MNRAS, 483, 276

\bibitem[\protect\citeauthoryear{Zjupa, Schaefer  \& Hahn}{Zjupa
  et~al.}{2020}]{Zjupa_tng_2020}
Zjupa J.,  Schaefer B.~M.,   Hahn O.,  2020, to be submitted to MNRAS

\bibitem[\protect\citeauthoryear{de Jong, Verdoes~Kleijn, Kuijken  \&
  Valentijn}{de~Jong et~al.}{2013}]{de_jong_kilo-degree_2013}
de Jong J. T.~A.,  Verdoes~Kleijn G.~A.,  Kuijken K.~H.,   Valentijn E.~A.,
  2013, \mn@doi [Experimental Astronomy] {10.1007/s10686-012-9306-1}, 35, 25

\bibitem[\protect\citeauthoryear{de
  Vaucouleurs}{de~Vaucouleurs}{1948}]{de_vaucouleurs_recherches_1948}
de Vaucouleurs G.,  1948, Annales d'Astrophysique, 11, 247

\bibitem[\protect\citeauthoryear{van Waerbeke, Bernardeau  \& Mellier}{van
  Waerbeke et~al.}{1999}]{van_waerbeke_efficiency_1999}
van Waerbeke L.,  Bernardeau F.,   Mellier Y.,  1999, A+A, 342, 15

\makeatother
\end{thebibliography}

\bsp
\label{lastpage}
\end{document}